\documentclass[12pt,epsf,amstex]{iopart}

\usepackage [dvips]{graphicx}
\usepackage{iopams}
\usepackage{amssymb}
\usepackage{amsthm}
\usepackage{epsfig}
\usepackage{verbatim}
\usepackage{color}

\definecolor{refkey}{rgb}{0,0.5,0.5}
\definecolor{labelkey}{rgb}{0.5,0,0.5}
\definecolor{orange}{rgb}{1.0,0.75,0}

\addtocounter{secnumdepth}{1}
\font\capfont=cmbx12 at 50 pt 
\newbox\capbox \newcount\capl \def\a{A}
\def\docappar{\medbreak\noindent\setbox\capbox\hbox{
\capfont\a\hskip0.15em}\hangindent=\wd\capbox
\capl=\ht\capbox\divide\capl by\baselineskip\advance\capl by1
\hangafter=-\capl
\hbox{\vbox to8pt{\hbox to0pt{\hss\box\capbox}\vss}}}
\def\cappar{\afterassignment\docappar\noexpand\let\a }
\def\bluew#1{{\color{blue} #1}}
\def\orangew#1{{\color{orange} #1}}

\newtheorem{Rule}{Rule}

\begin{document}

\newcommand{\ee}{{\rm e}}
\newcommand{\dd}{{\rm d}}
\newcommand{\p}{\partial}
\newcommand{\phX}{\phantom{XX}}

\newcommand{\bc}{\mathbf{c}}
\newcommand{\bC}{\mathbf{C}}
\newcommand{\bex}{\mathbf{e}_x}
\newcommand{\bey}{\mathbf{e}_y}
\newcommand{\bexy}{\mathbf{e}_{x,y}}
\newcommand{\bq}{\mathbf{q}}
\newcommand{\brr}{\mathbf{r}}
\newcommand{\bv}{\mathbf{v}}

\newcommand{\symbp}{{\rm E}}
\newcommand{\symbm}{{\rm N}}

\newcommand{\pE}{\symbp}
\newcommand{\pN}{\symbm}

\newcommand{\pEN}{\pE,\pN}
\newcommand{\pNE}{\pN,\pE}
\newcommand{\rhobar}{\overline{\rho}}
\newcommand{\rhop }{\rho^{\pE}}
\newcommand{\rhom }{\rho^{\pN}}
\newcommand{\rhopm}{\rho^{\pEN}}
\newcommand{\rhomp}{\rho^{\pNE}}
\newcommand{\rhobarp}{\overline{\rho}^{\pE}}
\newcommand{\rhobarm}{\overline{\rho}^{\pN}}
\newcommand{\rhobarpm}{\overline{\rho}^{\pEN}}
\newcommand{\etabar}{\overline{\eta}}
\newcommand{\etabarp}{\overline{\eta}^{\pE}}
\newcommand{\etabarm}{\overline{\eta}^{\pN}}
\newcommand{\etap}{\eta^{\pE}}
\newcommand{\etam}{\eta^{\pN}}
\newcommand{\etapm}{\eta^{\pEN}}
\newcommand{\drhop }{\delta\rho^{\pE}}
\newcommand{\drhom }{\delta\rho^{\pN}}
\newcommand{\drhopm}{\delta\rho^{\pEN}}
\newcommand{\drhomp}{\delta\rho^{\pNE}}
\newcommand{\hatrhop}{\hat{\rho}^{\pE}}
\newcommand{\hatrhom}{\hat{\rho}^{\pN}}
\newcommand{\hatrhopm}{\hat{\rho}^{\pEN}}
\newcommand{\hatdrhop}{\delta\hat{\rho}^{\pE}}
\newcommand{\hatdrhom}{\delta\hat{\rho}^{\pN}}
\newcommand{\hatdrhopm}{\delta\hat{\rho}^{\pEN}}
\newcommand{\tilderhop}{\tilde{\rho}^{\pE}}
\newcommand{\tilderhom}{\tilde{\rho}^{\pN}}
\newcommand{\tilderhopm}{\tilde{\rho}^{\pEN}}
\newcommand{\vp}{v^{\pE}}
\newcommand{\vm}{v^{\pN}}
\newcommand{\vpm}{v^{\pEN}}
\newcommand{\np}{n^{\pE}}
\newcommand{\nm}{n^{\pN}}
\newcommand{\npm}{n^{\pEN}}
\newcommand{\Jp}{J^{\pE}}
\newcommand{\Jm}{J^{\pN}}
\newcommand{\Jpm}{J^{\pEN}}

\newcommand{\alphac}{\alpha^{\rm c}}
\newcommand{\alphaM}{\alpha^{M}}
\newcommand{\alphaMm}{{\alpha^{M}_m}}
\newcommand{\alphaten}{\alpha^{10}}
\newcommand{\betaM}{\beta^{M}}
\newcommand{\betaMm}{{\beta^{M}_m}}

\newcommand{\hh}{\frac{1}{2}}
\newcommand{\la}{\langle}
\newcommand{\ra}{\rangle}
\newcommand{\beq}{\begin{equation}}
\newcommand{\eeq}{\end{equation}}
\newcommand{\bea}{\begin{eqnarray}}
\newcommand{\eea}{\end{eqnarray}}
\def\lsim{\:\raisebox{-0.5ex}{$\stackrel{\textstyle<}{\sim}$}\:}
\def\gsim{\:\raisebox{-0.5ex}{$\stackrel{\textstyle>}{\sim}$}\:}

\title[Crossing pedestrian traffic flows,diagonal stripe pattern, and chevron effect]{Crossing pedestrian traffic flows,diagonal stripe pattern, and chevron effect}

\author{J. Cividini$^1$, H.J.~Hilhorst$^2$, and C. Appert-Rolland$^3$}

\address{Laboratoire de Physique Th\'eorique, b\^atiment 210
Universit\'e Paris-Sud and CNRS (UMR 8627),
91405 Orsay Cedex, France}
\ead{$^1$~julien.cividini@th.u-psud.fr, $^2$~henk.hilhorst@th.u-psud.fr, $^3$~cecile.appert-rolland@th.u-psud.fr}

\begin{abstract}
We study two perpendicular intersecting 
flows of pedestrians. The latter are
represented either by moving hard core particles
of two types, eastbound ($\symbp$) and northbound ($\symbm$),
or by two density fields, $\rhop_t(\brr)$ and $\rhom_t(\brr)$.
Each flow takes place on a lattice strip of width $M$ so that the
intersection is an  $M\times M$ square.
We investigate the spontaneous formation, observed experimentally and in
simulations, of a diagonal 
pattern of stripes in which alternatingly one of the two 
particle types dominates. 
By a linear stability analysis of the field
equations we show how this pattern formation comes about.
We focus on the observation, reported recently,
that the striped pattern actually consists of chevrons rather than 
straight lines.
We demonstrate that this `chevron effect' occurs both in particle simulations 
with various different update schemes and in field simulations. 
We quantify the effect in terms of the chevron angle $\Delta\theta_0$
and determine its dependency on the 
parameters governing the boundary conditions.  
\end{abstract}


\maketitle

%%%%%%%%%%%%%%%%%%%%%%%%%%%%%%%%%%%%%%%%%%%%%%%%%%%%%%%%%%%%%%%%%%%%%%%%%%%%%
%%%%%%%%%%%%%%%%%%%%%%%%%%%%%%%%%%%%%%%%%%%%%%%%%%%%%%%%%%%%%%%%%%%%%%%%%%%%%
%%%%%%%%%%%%%%%%%%%%%%%%%%%%%%%%%%%%%%%%%%%%%%%%%%%%%%%%%%%%%%%%%%%%%%%%%%%%%

\section{Introduction} 
\label{sect_introduction}

%%%%%%%%%%%%%%%%%%%%%%%%%%%%%%%%%%%%%%%%%%%%%%%%%%%%%%%%%%%%%%%%%%%%%%%%%%%%%

\subsection{General}
\label{section:introgeneral}

\cappar The common feature between the physical systems
of traditional statistical mechanics and traffic models 
is that both deal with 
interacting entities engaged in collective motion: 
atoms or molecules in the case of the physics
of fluids, macroscopic particles
in the case of flowing granular matter,
and cars or pedestrians in the case of road traffic \cite{schadschneider2008b,helbing2001b}.
This analogy explains the physicists' interest in traffic models.
One approach is to develop traffic models that render the traffic flow
as accurately as possible in concrete situations, whether it be
entrance or exit ramps of highways, traffic lights at intersections,
or others.
This `specific' approach, useful and necessary for 
real-life traffic control problems,
is complemented by a more theoretical one in which one
attempts to extract the most general features of a large variety of
traffic situations. 
These features may then be looked for also in new situations. 
Simplified models may in particular evidence basic mechanisms leading to pattern formation \cite{cross_h1993,wolfram1983,packard_w1985}.
This approach is of course strongly influenced by the physicists'
experience with critical phenomena, where universal properties
are at the heart of the theory.
An important task then is to try to find classes of
traffic models similar to the universality classes of
critical phenomena \cite{wolfram1984}. Such an effort, if successful, would greatly
structure the body of knowledge in this relatively young field of research.
In the present work we take this approach.
\vspace{2mm}

We are interested here in the problem of two crossing traffic flows.
Crossing flows, whether consisting of pedestrians or vehicles, 
have been the subject of a great number of studies.
Some of them deal with the crossing of two single lanes,
as for example in references 
\cite{nagel_s1992,fouladvand_s_s04b,foulaadvand_n07,du2010,appert-rolland_c_h2011c}.
Of interest in this paper are the crossings of wider streets. 
The intersection of two perpendicular one-way streets of width $M$ 
is a square $M\times M$ domain that we will refer to as the
{\it intersection square}.
In the literature the modeling of traffic flows
on such intersecting streets has taken various forms.
One class of models is based on describing the
motion of individual particles, and another one
on replacing each particle species by a space- and time dependent 
density field.
The particle motion may be either in continuous space or on a lattice,
and similarly the fields may be defined in continuous space
or on a lattice. 

Below we briefly mention the existing work most
relevant to ours.

%%%%%%%%%%%%%%%%%%%%%%%%%%%%%%%%%%%%%%%%%%%%%%%%%%%%%%%%%%%%%%%%%%%%%%%%%%%%%

\subsection{Stripe formation and chevron effect}
\label{section:introstripe}

\begin{figure}
\begin{center}
\scalebox{.35}
{\includegraphics{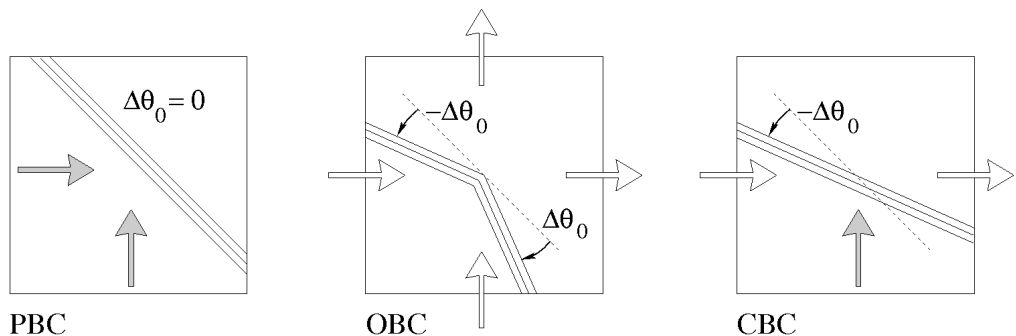}} 
\end{center}
\caption{\small Schematic representation of the
intersection square of two perpendicular traffic flows for three
different types of boundary conditions: periodic (PBC), open (OBC),
and cylindrical (CBC). White arrows correspond to incoming and
outgoing fluxes, grey arrows to fluxes that circulate with periodic
boundaries. 
The dashed lines are for reference and are
at $45^\circ$ with respect to the main axes. 
Solid lines indicate the average
orientation of the striped pattern; the angle
differences $\pm\Delta\theta_0$ with respect to the line of reference
have been exaggerated for greater clarity.
For OBC the pattern is chevron-like, for CBC only one
branch of the chevron persists.
}   
\label{fig_stripesbc}
\end{figure}

The model defined in 1993 by Biham, Middleton and Levine \cite{biham_m_l1992}
(the `BML model') has come to enjoy a 
definite popularity. This model,
whose first aim was to describe urban traffic in a Manhattan-like city,
is a deterministic cellular automaton that
deals not with actually crossing streets but with the perpendicular
motion of two particle species on a torus, that is,
a square $M\times M$  lattice  with periodic boundary conditions.
Each lattice site may be occupied
by either an eastbound or a northbound particle, or be empty.
The update is parallel but alternating between 
the two particle species.
The authors observed that for low enough values of the two densities 
(equal to a common value $\rho$)
the particles organize into a {\it pattern of diagonal stripes\,}
at an angle of $45^{\circ}$ to both flow directions 
(see Fig.\,\ref{fig_stripesbc}a), 
each stripe exclusively containing particles of one of the two kinds
(reference \cite{biham_m_l1992}, Fig.\,1,
which is for $M=32$ and $\rho=0.125$).
Biham {\it et al.} remark that this arrangement allows the particles to
achieve their maximum speed, which is close or equal to
one lattice distance per time step.

Recently however, Ding {\it et al.} \cite{ding_j_w2011,ding2011}, after
replacing the alternating parallel update of the BML model by a
random sequential update, no longer observed this diagonal pattern 
(Fig.\,2 in Ref.\,\cite{ding_j_w2011}, which is for
$M=100$ and $\rho=0.10$).

Hoogendoorn and Bovy \cite{hoogendoorn_b2003}
modeled interacting walkers in continuum space
using a cost function that is to be minimized and that
incorporates parameter values drawn from real-life situations.
When applied to walkers in crossing hallways, this model again exhibits
the formation of striped patterns 
(Fig.\,4 in Ref.\,\cite{hoogendoorn_b2003}, where the width of the
hallways is of the order of a few meters). 
The authors consider this phenomenon as being in the same class 
as that of lane formation in bidirectional flow
\cite{moussaid2012}, also shown in their work 
(Fig.\,3 in Ref.\,\cite{hoogendoorn_b2003}).
Lane formation is of course
well-known in a class of nonequilibrium models of statistical mechanics
\cite{schmittmann_z1995,dzubiella_h_l2002}.

Yamamoto and Okada \cite{yamamoto_o2011}
simulated crossing pedestrian flow with the purpose
of conceiving control methods rendering such flows 
smoother and safer.
They determine the velocity vector of each particle by means of 
an auxiliary field that takes into account both the particle's
destiny (north or east) and the proximity of other particles.
This model again reproduces the striped pattern in the
intersection area (which in this case is only approximately square,
Fig.\,3 in Ref.\,\cite{yamamoto_o2011}).
The authors then go on and
convert this particle model into 
one where each of the two particle species is
replaced with a space and time dependent density field.
The field equations are essentially of the mean field type, that is,
each particle interacts no longer with the
other particles individually, but with their densities. 

\vspace{2mm}

Stripe formation thus appears as a phenomenon that is common to 
a wide class of crossing flow models, the criterion
apparently being that the flows are unidirectional and sufficiently
deterministic.
The stripes appear in a density regime (the `free flow phase') between zero
and some upper critical value above which the intersection
square undergoes jamming.
In this work we provide further evidence for the ubiquity
of stripe formation by studying it in two types of models,
both defined on a lattice. The first one
is the particle model introduced in Ref.\,\cite{hilhorst_a2012},
and the second one is a mean field version of it.
The intersection square is an open system with 
flows that enter and exit; however, in addition to these open boundary
conditions (OBC) we will be led, in analogy with the BML model,
also to consider interaction squares subject to
periodic boundary conditions in one or in both directions,
to which we will refer as cylindrical (CBC) and periodic (PBC) 
boundary conditions, respectively.

The remainder of this work focuses on a new phenomenon that accompanies
the stripe formation under OBC and that is
called the {\it chevron effect}; its discovery was
first reported in Ref.\,\cite{cividini_a_h2013}.
It is the subtle phenomenon that the stripes deviate from straight lines
but are actually chevrons, that is, each stripe
consist of two straight lines at angles\footnote
{Throughout this paper, angles of straight lines 
are measured clockwise from the west.\label{foot1}}
of $45^{\circ}\pm \Delta\theta_0$ that join 
on the symmetry axis (see Fig.\,\ref{fig_stripesbc}b). 
Throughout this paper, angles of straight lines 
are measured clockwise from the west.
The angle difference
$\Delta\theta_0$, which is of the order of a degree,
is referred to as the {\it chevron angle}.
We show that the chevron effect, too, occurs in both the particle and
the nonlinear mean field model. The manifestation of the effect
is boundary condition dependent.
Under CBC only `half' of the chevron effect subsists:
there appears only a single branch of the chevron,
but it still has the same angle difference $\Delta\theta_0$ 
with respect to the diagonal (see Fig.\,\ref{fig_stripesbc}c).
\vspace{3mm}

%%%%%%%%%%%%%%%%%%%%%%%%%%%%%%%%%%%
%%%%%%%%%%%%%%%%%%%%%%%%%%%%%%%%%%%
%%%%%%%%%%%%%%%%%%%%%%%%%%%%%%%%%%%
%%%%%%%%%%%%%%%%%%%%%%%%%%%%%%%%%%%

This paper is organized as follows.
In section \ref{section:micro} we describe the particle model.
In section \ref{section:meanfield}
we relate it to a corresponding mean-field model.
In section \ref{section:pbc} we simulate the particle system with PBC.
We exhibit the instability by which a uniform initial
state develops into a stationary state with a diagonal pattern
of stripes.  
We explain this phenomenon in terms of a linear stability analysis
of the mean field model. 
In section \ref{section:obc} we simulate the same stripe formation
instability with OBC.
We point out the chevron effect which, whereas absent for PBC,
appears under OBC both in the particle and the mean field model.
We discuss two methods of measuring and quantifying the 
chevron angle $\Delta\theta_0$,
the `crest method' and the `velocity ratio method', and
show by simulation that $\Delta\theta_0$ is linear in the particle
density. 
In section \ref{section:chevrontheory} 
we provide an elementary theoretical argument that
explains the chevron effect.
In section \ref{section:cbc} we simulate the mean field model with
CBC. In this case there is a control parameter for each of the two
directions (basically the densities of the two particle fluxes),
and we determine how $\Delta\theta_0$ depends on them.
Whereas all the above work deals with stationary states,
in section \ref{secchevrontransient}
we shed additional light on the chevron effect 
by studying it in a transient. In section \ref{section:conclusion} 
we present a summary of our results and conclude.

%%%%%%%%%%%%%%%%%%%%%%%%%%%%%%%%%%%%%%%%%%%%%%%%%%%%%%%%%%%%%%%%%%%%%%%%%%%%

\section{Particle model}
\label{section:micro}

%%%%%%%%%%%%%%%%%%%%%%%%%%%%%%%%%%%%%%%%%%%%%%%%%%%%%%%%%%%%%%%%%%%%%%%%%%%%

\subsection{Geometry}
\label{section:geometry}

\begin{figure}
\begin{center}
\scalebox{.55}
{\includegraphics{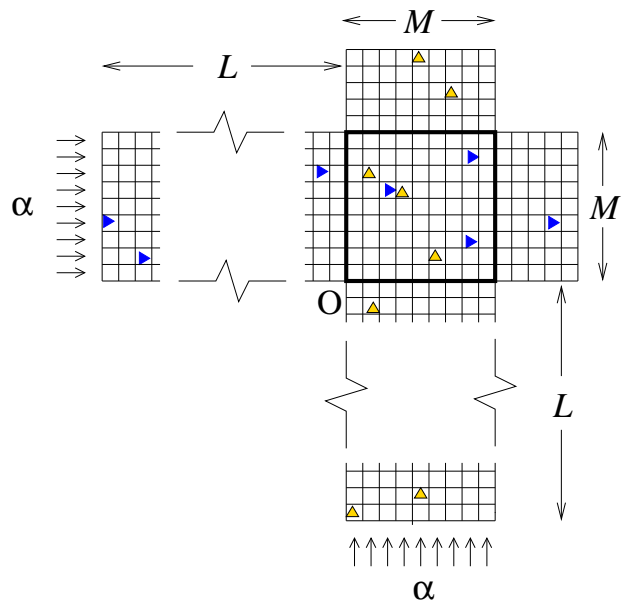}} 
\end{center}
\caption{\small The intersection of two streets of width $M$ is the
  $M\times M$ square lattice surrounded in bold.
  Eastbound particles are represented by blue
  triangles pointing right and northbound ones by orange
  ones pointing upward. The symbol `{\bf O}' denotes the origin of the
  coordinate system; the site in the lower left corner of the square
  has coordinates $(1,1)$.
  The influx of particles takes place at a large distance $L$ 
  from the intersection square and is controlled by a parameter $\alpha$.
}  
\label{fig_intersectinglanes}
\end{figure}

A `street of width $M$' is modeled on a lattice as a set
of $M$ parallel infinite one-dimensional lanes.
Two such streets intersecting at a right angle lead to the geometry of 
Fig.\,\ref{fig_intersectinglanes}. The heavy line surrounds
the $M\times M$ intersection square, whose sites we will denote by
$\brr=(i,j)$ with $1\leq i, j\leq M$. 
At a large distance $L$ from the intersection square,
eastbound $(\pE) $ and northbound $(\pN)$ particles
are injected onto the `injection sites' (the ones indicated by arrows
in Fig.\,\ref{fig_intersectinglanes}) of the
the horizontal and vertical lanes, respectively,
with a probability $\alpha$ 
per unit time interval for each empty injection site.
Each particle stays in its lane and advances by steps of a single
lattice unit according to an update scheme to be discussed below. 
The $(1,1)$ diagonal through the origin is, statistically, 
an axis of symmetry.
The geometry described here
was introduced and studied in Ref.\,\cite{hilhorst_a2012};
there, however, the focus was on the jamming
transitions that occur when $\alpha$ exceeds
a critical value $\alpha_{\rm c}(M)$, which
for the $M$ values considered in this work is
typically of the order of $0.10$ .
In the present study we will consider the model in the regime of 
low $\alpha$.
We will employ two distinct update schemes. 

%%%%%%%%%%%%%%%%%%%%%%%%%%%%%%%%%%%
%%%%%%%%%%%%%%%%%%%%%%%%%%%%%%%%%%%
%%%%%%%%%%%%%%%%%%%%%%%%%%%%%%%%%%%
%%%%%%%%%%%%%%%%%%%%%%%%%%%%%%%%%%%

%%%%%%%%%%%%%%%%%%%%%%%%%%%%%%%%%%%%%%%%%%%%%%%%%%%%%%%%%%%%%%%%%%%%%%%%%%%%%

\subsection{Update schemes}
\label{secupdateschemes}
 
The first update scheme that we will use is the {\it frozen shuffle update}
\cite{appert-rolland_c_h2011a,appert-rolland_c_h2011b,appert-rolland_c_h2011c}.
Under this scheme, each particle $\ell$ that enters the system
is assigned a phase 
$\tau_\ell\in[0,1)$ \cite{appert-rolland_c_h2011b},
which it keeps until it leaves the system again; 
its position is then updated at the
instants $t+\tau_\ell$ on the continuous time axis, 
where $t$ is an integer. 
An update consists in
moving the particle one site ahead unless its target site is occupied.
Hence during a unit interval all particles positions are updated once, 
and this happens in order of increasing phases (the `update sequence').

This update is suitable for pedestrians because it 
reproduces step cycles with a distribution of phases.

The second update scheme that we will use is the 
{\it alternating parallel update}, 
in which the $\symbp$ particles are updated in parallel at 
half-integer times
and the $\symbm$ particles all in parallel at integer times. 
An update consists in simultaneously 
moving all those particles whose target site is empty.

In two-dimensional situations such as occur on the
intersection square, both schemes have the advantage of 
avoiding `conflicts', that is, of providing a
natural priority rule when two perpendicularly traveling particles
target the same site. 
At the relatively low particle densities 
that we are concerned with here, our simulations will show
that many features of the behaviour of the system
are qualitatively the same for both update schemes.

%%%%%%%%%%%%%%%%%%%%%%%%%%%%%%%%%%%%%%%%%%%%%%%%%%%%%%%%%%%%%%%%%%%%%%%%%%%%%

\subsection{Open-ended boundary conditions}

In the street sections of length $L$ waiting lines
may form at the entrance of the intersection square.
For $\alpha<\alpha_{\rm c}(M)$ this waiting line has
a fluctuating length of some finite stationary average value
and we will say that the system is in the {\it free flow phase}.
The opposite case, $\alpha>\alpha_{\rm c}(M)$, 
leads to a {\it jammed phase} 
and will not concern us here. 
In the free flow phase
the control parameter $\alpha$ entirely
determines the average injected current $J(\alpha)$, which
is also equal to the current passing through each lane.
The expressions are
\beq
J(\alpha)=\left\{
\begin{array}{ll}
\alpha/(1+\alpha), & \mbox{alternating parallel update},\\[2mm]
a/(1+a), & \mbox{frozen shuffle update}.
\end{array}
\right.
\label{xJ}
\eeq
in which $a\equiv-\log(1-\alpha)$ is the injection {\it rate\,}
corresponding to $\alpha$.
The first one of relations (\ref{xJ}) is well-known and the second one was 
derived in Ref.\,\cite{appert-rolland_c_h2011b}.
Both lead to $J(\alpha)\simeq\alpha$ in the small $\alpha$
(low density) limit.

In practice we chose $L$ (see
Fig.\,\ref{fig_intersectinglanes}) larger than the 
lengths of any waiting lines observed in the simulations,
so that effectively $L=\infty$.\footnote{ 
In Ref.\,\cite{hilhorst_a2012} it was shown how to reduce a simulation
in which the waiting lines may become arbitrarily long to a simulation on
the intersection square involving only a finite number of variables.
This method is indispensable for the study of the jamming transition
but was not used here.}
We therefore have a two-parameter model whose properties
depend only on $\alpha$ and $M$. 

%%%%%%%%%%%%%%%%%%%%%%%%%%%%%%%%%%%%%%%%%%%%%%%%%%%%%%%%%%%%%%%%%%%%%%%%%%%%%%

\subsection{Equations for the particle model}
\label{section:particlemodel}

For both update schemes
the system evolves like a deterministic cellular automaton
with stochastic boundary conditions. 
Let $n^{\pE}_{s}(\brr)=1$ (or $=0$) if at time $s$ site $\brr=(i,j)$ is (or is
not) occupied by an eastbound particle. A similar definition holds for $n^{\pN}_{s}(\brr)$. 
The update scheme (with $s$ continuous or discrete, as the case may be) and the boundary conditions together
determine the time evolution of these occupation numbers.
For the case of alternating parallel update the occupation
numbers of the eastbound particles satisfy 
\beq
n^{\pE}_{t+\hh}(\brr) = [ 1-n^{\pE}_{t-\hh}(\brr)-n^{\pN}_{t}(\brr) ]
                      n^{\pE}_{t-\hh}(\brr-\bex)
                     + [ n^{\pE}_{t-\hh}(\brr+\bex)+n^{\pN}_{t}(\brr+\bex) ]
                      n^{\pE}_{t-\hh}(\brr),
\label{xoccnumbers}
\eeq
for any integer $t$ and where
$\bex$ is the unit vector along the $i$ direction;
and those for the northbound particles satisfy an analogous equation
relating their values at the integer times $t+1$ to those at $t$. 
For frozen shuffle update described in section \ref{secupdateschemes}, in contrast to the simplicity of the
numerical algorithm, the analytic expression for the time
evolution equations  
is quite cumbersome and we will not display it\footnote
{It depends on the full set of parameters $\{\tau_\ell\}$,
where $\ell$ runs through all particles present in the system at the
instant of time under consideration.}.   

Of interest are, for each update scheme, the mean values 
$\langle n^{\pEN}_t(\brr)\rangle$, where 
the average $\langle\ldots\rangle$
is over the stochastic boundary conditions at the injection sites
and possibly over stochastic initial conditions at time $t=0$.

%%%%%%%%%%%%%%%%%%%%%%%%%%%%%%%%%%%%%%%%%%%%%%%%%%%%%%%%%%%%%%%%%%%%%%%%%%%%%%

\section{Mean field model}
\label{section:meanfield}

%%%%%%%%%%%%%%%%%%%%%%%%%%%%%%%%%%%%%%%%%%%%%%%%%%%%%%%%%%%%%%%%%%%%%%%%%%%%%%

\subsection{Equations for the mean field model}
\label{section:eqnsmf}

In this section we will juxtapose
the particle model formulated in terms of the binary variables
$n^{\pE}_t(\brr)$ and $n^{\pN}_t(\brr)$ with a mean field description in
terms of continuous fields $\rhop_t(\brr)$ and $\rhom_t(\brr)$,
the latter being thought of as representing local averages. 
With the purpose of retaining only
the strict minimum of terms
we introduce one further approximation, that consists in
neglecting in (\ref{xoccnumbers}) the interactions between particles 
of the same kind. {\it A partial} justification for this runs as follows.
Let $\rho$ be the typical density of each of the two particle types in
the intersection square.
In the low density limit
the positions of the E particles are not correlated with those
of the N particles, and hence 
the frequency of a local blocking event between two particles
of different types tends to zero as the square of their density, $\sim \rho^2$. 
On the contrary, two consecutive same-type particles in the same lane 
(a `leader' and a `follower', say of type E) 
have their positions and speeds correlated in such a way that the leader cannot
block the follower unless it is first blocked itself by an
N particle; and the frequency of such a local three particle event
is proportional to $\rho^3$. 

When we average equations (\ref{xoccnumbers}), correlations between
occupation numbers appear. We obtain a closed system of equations
through the mean field approximation that consists in
factorizing these correlations, that is, we
simply repace in (\ref{xoccnumbers}) the $n^{\pEN}_t(\brr)$ by 
their averages $\rhopm_t(\brr)$. If moreover we pass
from alternating parallel to fully parallel update, we get
\bea
\label{mfeqns2d}
\rhop_{t+1}(\brr) = [1 - \rhom_t(\brr)]\rhop_t(\brr-\bex) + \rhom_t(\brr+\bex)\rhop_t(\brr),\\
\rhom_{t+1}(\brr) = [1 - \rhop_t(\brr)]\rhom_t(\brr-\bey) + \rhop_t(\brr+\bey)\rhom_t(\brr). \nonumber
\eea
These equations define what we will refer to as the mean field model.
It is hard to assess {\it a priori\,} the quality of the approximation
involved in going from the particle model to (\ref{mfeqns2d}),
whatever the update be. In the following sections we will 
solve Eqs.\,(\ref{mfeqns2d}) numerically under a variety of
boundary conditions and observe that the behavior of the 
densities $\rhopm_t(\brr)$ is qualitatively close to that of the 
particle densities in the particle model. 
We will then take this correspondence
as an {\it a posteriori\,} confirmation that 
Eqs.\,(\ref{mfeqns2d}) make sense.
\vspace{2mm}

{\it Linearized equations.\,\,} 
We notice that a uniform density distribution 
$\rhopm_t(\brr) = \rhobar$\, solves Eqs.\,(\ref{mfeqns2d}).
Setting $\rhopm_t(\brr) = \rhobar + \drhopm_t(\brr)$ 
and linearizing these equations in $ \drhopm_t$ one obtains 
\bea 
\drhop_{t+1}(\brr) &= (1 - \rhobar)\drhop_t(\brr-\bex) -
\rhobar\drhom_t(\brr) + \rhobar\drhop_t(\brr) +
\rhobar\drhom_t(\brr+\bex),\nonumber\\[2mm]
\drhom_{t+1}(\brr) &= (1 - \rhobar)\drhom_t(\brr-\bey) -
\rhobar\drhop_t(\brr) + \rhobar\drhom_t(\brr) +
\rhobar\drhop_t(\brr+\bey).\nonumber\\ 
&
\label{mfeqns2dlin}
\eea
In this study we will, on the one hand, perform simulations of
the particle model, and on the other hand
investigate the numerical solution of the mean 
field model (\ref{mfeqns2d}). We will briefly mention some
analytic work that may be done on (\ref{mfeqns2dlin}).

%%%%%%%%%%%%%%%%%%%%%%%%%%%%%%%%%%%%%%%%%%%%%%%%%%%%%%%%%%%%%%%%%%%%%%%%%%%%

\subsection{Open-ended boundary conditions}
\label{secbcmf}

For the mean-field equations (\ref{mfeqns2d}) we implement the open-ended
boundary conditions (OBC)
for simplicity in a way slightly different from how
we applied them to the particle system.
We will in fact restrict the simulation to the intersection square,
that is, to the lattice sites $\brr=(i,j)$ with $1\leq i,j\leq M$.
Equations (\ref{mfeqns2d}) couple 
the boundary site  $\brr=(1,j)$ to the site
$\brr-\bex = (0,j)$ outside this square. 
We will refer to the sites $(0,j)$ and $(i,0)$ as the 
{\it entrance sites} at the west and south boundary, respectively, 
of the intersection square.
Instead of imposing the injection rate $\alpha$ a large distance $L$ away
from the intersection square, as we did for the
particle model, we will suppress the street segments of length $L$
leading up to the square
and impose at each instant of time $t$ the random entrance site
densities 
\beq
\rhop_t(0,k) = \etap_t(k),  \qquad \rhom_t(k,0) = \etam_t(k), 
\label{obcond}
\eeq
for all $k=1,2,\ldots,M$, where the $\etapm_t(k)$ are i.i.d. 
random variables $\eta$ of average $\etabar$. 
In actual practice, supposing that the details of their
probability law are unimportant, we drew them from the uniform
distribution 
\beq
p(\eta;\etabar) = 
\left\{
\begin{array}{ll}
{1}/{\etabar},  & 
\frac{1}{2}\etabar<\eta<\frac{3}{2}\etabar,\\[2mm]
0, & \mbox{else}.
\end{array}
\right.
\label{dprho}
\eeq
With the random boundary conditions (\ref{obcond})
the $\rhopm_t(\brr)$ become
random variables and we will indicate their averages
by $\la\rhopm_t(\brr)\ra$. 

As exit boundary conditions we take
$\rhop_t(M+1,k) = \rhom_t(k,M+1) = 0$, which expresses
that the particles freely leave the system.
Note that when complemented with these exit boundary conditions,  Eqs.(\ref{mfeqns2d}) are not solved anymore by a uniform density for a finite system but lead to a boundary effect at the exit. Thus Eqs.(\ref{mfeqns2dlin}) can rigorously be regarded as the linearization of Eqs.(\ref{mfeqns2d}) only in the $M \rightarrow \infty$ limit.

With these free exit conditions, the currents through each
of the $2M$ lanes are determined by the entrance boundary conditions. As these are expressed in Eq.(\ref{obcond}) as {\it density boundary conditions}, we must expect
the currents $J_k(\etabar)$ to depend on the lane index $k$;
this dependency, however, has turned out to be extremely weak 
in all situations studied in this work.

Finally, whereas the mean field model of basic interest has the open boundary
conditions described above,
we will also be led, in the sections 
below to replace these with periodic boundary
conditions in one or both of the directions.

%%%%%%%%%%%%%%%%%%%%%%%%%%%%%%%%%%%%%%%%%%%%%%%%%%%%%%%%%%%%%%%%%%%%%%%%%%%%%

\section{Crossing flows on a torus}
\label{section:pbc}

In order to understand the basic mechanism of the
stripe formation we first study, in this section, 
a simplified problem in which  
the interaction square is submitted to
periodic boundary conditions (PBC) in both
directions. This procedure was first proposed by Biham {\it et al.}
\cite{biham_m_l1992} and later followed by several other authors
\cite{ding_j_w2011}. The results from such a study on a torus are thus
of interest in their own right. Our investigation includes an analytic
calculation which explains the instability observed on the torus in
our own and in earlier simulations.
Moreover, our PBC results will
serve as a basis for comparison 
when in the next section we study the original
problem, that is, the intersection square with open
boundaries (OBC). 

%%%%%%%%%%%%%%%%%%%%%%%%%%%%%%%%%%%%%%%%%%%%%%%%%%%%%%%%%%%%%%%%%%%%%%%%%%%%%

\subsection{Stripe formation in the particle model}
\label{sec:simulationspbc}

We consider a particle system
with frozen shuffle update and
impose on the intersection square PBC in both directions of space.
At the initial time $2N$ particles ($N$ going east and $N$ going
north) are placed on random lattice positions subject to hard core
exclusion. For these boundary conditions the
space averaged particle density $\rhobar=N/M^2$
replaces $\alpha$ as the control parameter. 
Choosing the phases $\{\tau_\ell\}$ of a conserved set of particles
amounts to choosing a fixed random permutation of them.
They are then updated at each time step in that order.
The system therefore is a deterministic cellular automaton:
its initial state determines, {\it via\,} the update scheme, 
its entire time evolution.

%%%%%%%%%%%%%%%%%%%%%%%%%%%%%%%%%%%%
%%%%%%%%%%%%%%%%%%%%%%%%%%%%%%%%%%%%
\begin{figure}
\begin{center}
\scalebox{.4}
{\includegraphics{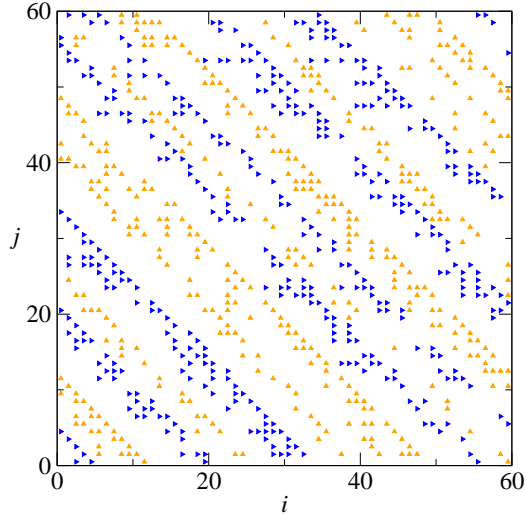}}
\end{center}
\caption{\small Snapshot of the stationary state of a particle
  simulation with frozen shuffle update and
periodic boundary conditions, for system size $M=60$ and particle
density $\rhobar=0.1$. Particles
of the same type are aligned along diagonals
in the $(1,-1)$ direction that have a width of a few lattice distances.
The simulation was carried out by Chlo\'e Barr\'e, who also
  prepared this figure.
} 
\label{fig_chloe}
\end{figure}
%%%%%%%%%%%%%%%%%%%%%%%%%%%%%%%%%%%%
%%%%%%%%%%%%%%%%%%%%%%%%%%%%%%%%%%%%

Simulations show that after a certain transient time $t_{\rm trans}$
the uniform particle distribution becomes unstable. 
Fig.\,\ref{fig_chloe} shows what happens for the example of a street
width $M=60$ containing 720 particles: the system
self-organizes into a pattern of alternating diagonals of same-type
particles that are at an angle
of $45^{\circ}$. 
The wavelength of the pattern is typically in the range from 5 to 15
lattice distances; it is irregular, its details depending on the
initial condition.
The system eventually enters a limit cycle in which all or almost all 
particles move at each time step. 

In the limit of small $\rhobar$ limit we found that
roughly $t_{\rm trans} \sim 1/\rhobar$.
For the linear lattice size $M=60$ (measured in lattice units) 
and the low particle densities $0.02\lesssim \rhobar\lesssim 0.10$ 
that we considered, we found that
$t_{\rm trans}$ (measured in time steps) 
may be up to an order of 
magnitude larger than the lattice size $M$.
This sets a limit to what we can learn from this PBC study
about the behavior of the open system:
if $t_{\rm trans} \gtrsim M$, the particles entering the open
intersection square will not be able to fully develop their
instability before they quit the system again. We will return to this
point in section \ref{section:obc}.

%%%%%%%%%%%%%%%%%%%%%%%%%%%%%%%%%%%%%%%%%%%%%%%%%%%%%%%%%%%%%%%%%%%%%%%%%%%%%%

\subsection{Stripe formation instability of the mean field model}
\label{secinstability1}

\begin{figure}
\begin{center}
\scalebox{.35}
{\includegraphics{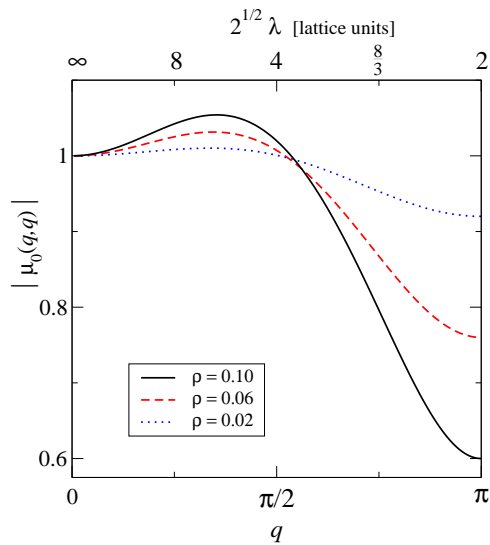}}
\end{center}
\caption{\small Larger absolute eigenvalue $| \mu_0(\mathbf{q}) |$ along
 the diagonal as a function of the wavenumber $q = q_x = q_y $. The
 upper axis shows the corresponding wavelength $\lambda=2\pi/q$.
}
\label{fig_absmu0q}
\end{figure}

{\it Linear regime.\,\,} 
In order to explain the origin of the instability analytically,
we will now perform a stability analysis
on the linearized mean field equations (\ref{mfeqns2dlin})
with PBC. Let us define the Fourier transforms 
\beq
\hatrhopm_t(\bq) = \sum_{k=1}^M  \sum_{l=1}^M 
\ee^{{\rm i}\bq\cdot\brr}\,\drhopm_t(k,l) 
\label{FTrho}
\eeq
where $\bq=(q_x,q_y)$
with $q_{x,y}=2\pi\kappa_{x,y}/{M}$ and $\kappa_{x,y}=0,1,\ldots,M-1$.
The linearized equations then read
\beq
\left(
\begin{array}{r}
\hatdrhop_{t+1}(\mathbf{q})\\
\hatdrhom_{t+1}(\mathbf{q})
\end{array}
\right)
=
\left(
\begin{array}{cc}
\ee^{{\rm i}q_x}R_{q_x} & R_{q_x}-1 \\
R_{q_y}-1 & \ee^{{\rm i}q_y}R_{q_y} 
\end{array}
\right)
\left(
\begin{array}{r}
\hatdrhop_t(\mathbf{q})\\
\hatdrhom_t(\mathbf{q})
\end{array}
\right) ,
\label{lineqs2d}
\eeq
where $R_q = 1+\rhobar(\ee^{-{\rm i}q}-1)$. 
One of the eigenvalues of the $2 \times 2$ matrix in (\ref{lineqs2d}) 
is always inside the unit circle.
The other one, which we will call  $\mu_0(\mathbf{q})$, is given by
\beq
\mu_0(\mathbf{q}) = \frac{1}{2}\Big( \ee^{{\rm i}q_x}R_{q_x} +
\ee^{{\rm i}q_y}R_{q_y} + \big[ (\ee^{{\rm i}q_x}R_{q_x}-\ee^{{\rm
    i}q_y}R_{q_y})^2 - 4 (R_{q_x} - 1) (R_{q_y} - 1) \big]^{1/2} \Big). 
\label{mu0}
\eeq
We found numerically that its absolute value $|\mu_0(\bq)|$ has
a maximum on the diagonal $q_x=q_y$, as shown in Fig.\,\ref{fig_absmu0q}. 
This maximum exceeds unity and is therefore associated with an unstable mode 
traveling in the $(1,1)$ direction with wavelength 
\bea
\lambda_{\rm max} = 2\pi/|\mathbf{q} |_{\rm max} &=& 
\sqrt{2} \pi/\arccos[(1-2\rhobar)/(2-2\rhobar)] \nonumber\\[2mm]
&=& 3 \sqrt{2}[1-(\sqrt{3}/\pi)\rhobar] + {\cal O}(\rhobar^2),
\label{xlambdamax}
\eea
where in the second line we expanded for small $\rhobar$.
This calculation thus explains the formation of a diagonal striped pattern as 
a consequence of the unstable Fourier modes present 
in a random initial state. 
\vspace{2mm}

%%%%%%%%%%%%%%%%%%%%%%%%%%%%%%%%%%%%
%%%%%%%%%%%%%%%%%%%%%%%%%%%%%%%%%%%%
%%%%%%%%%%%%%%%%%%%%%%%%%%%%%%%%%%%%
%%%%%%%%%%%%%%%%%%%%%%%%%%%%%%%%%%%%

{\it Nonlinear regime.\,\,} 
The nonlinear regime of the mean field equations, 
Eqs.\,(\ref{mfeqns2d}), is outside the reach of this analysis. 
Numerical solution of (\ref{mfeqns2d}) with PBC
shows that the solution tends
to a stationary state consisting of 
alternating stripes having on each site $\brr$
either $\rhop_t(\brr)\neq 0$ or
$\rhom_t(\brr)\neq 0$; and in which consecutive stripes are 
separated by unoccupied sites in such a way that all nonlinear
terms in (\ref{mfeqns2d}) vanish. 
The density patterns $\rhop_t(\brr)$ and $\rhom_t(\brr)$ 
then advance at unit speed unimpeded eastward and
northward, respectively, around the torus, 
in a way perfectly similar to the
particles in section \ref{sec:simulationspbc}.
In the final state under PBC, therefore, the coupling
between the two particle types has disappeared.

Armed with this understanding we now return
to the original problem of the open intersection square.

%%%%%%%%%%%%%%%%%%%%%%%%%%%%%%%%%%%%%%%%%%%%%%%%%%%%%%%%%%%%%%%%%%%%%%%%%%%%%

\section{Open intersection}
\label{section:obc}

We now address the central question of this work, that is, how do
two flows cross on an open square?
The control parameters are those defined in sections
\ref{section:micro} and \ref{section:meanfield},
namely the injection probability $\alpha$ in the
case of the particle model, and the boundary density $\etabar$
at the entrance sites in the case of the mean field model.
Whereas under PBC the final stationary state was determined 
by the randomly selected initial state, under OBC 
it is a consequence of the noise coming from the entrance
boundaries. In all the simulations and numerical calculations below
we took statistics only after a transient time sufficiently long for
the system to settle in a stationary state.
Typically, this time was of the order of a few times the linear
lattice size $M$.

%%%%%%%%%%%%%%%%%%%%%%%%%%%%%%%%%%%%%%%%%%%%%%%%%%%%%%%%%%%%%%%%%%%%%%%%%%%%%

\subsection{Stripe formation in the particle model}
\label{secinstability2}

\begin{figure}
\begin{center}
\scalebox{.35}
{\includegraphics{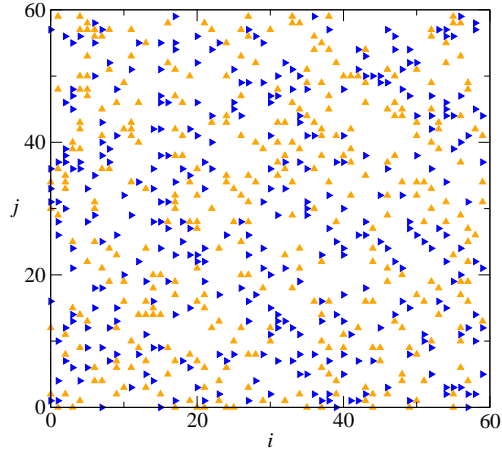}} 
\end{center}
\caption{\small Snapshot of the stationary state of the intersection
  square in a particle
  simulation with open boundary conditions,
  for $\alpha = 0.09$ and $M=60$. The blue particles
  (\bluew{$\blacktriangleright$}) come from the left and the orange ones
  (\orangew{$\blacktriangle$}) from the bottom. 
  Obtained with frozen shuffle update.}  
\label{fig_alpha60_0.09}
\end{figure}

\begin{figure}
\begin{center}
\scalebox{.36}
{\includegraphics{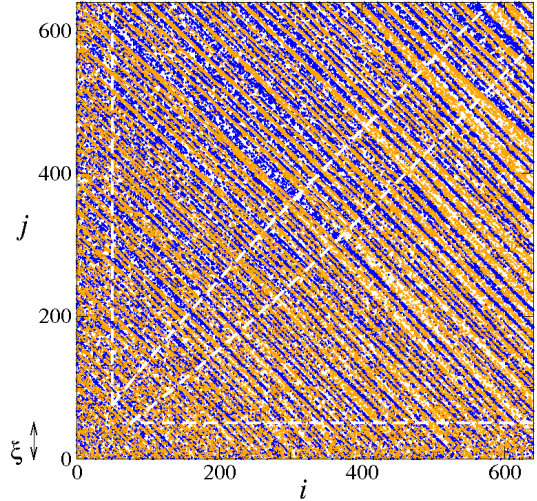}}
\end{center}
\caption{\small Snapshot of the stationary state of the intersection
  square of a particle system
  with $M = 640$ and $\alpha = 0.09$ and subject to frozen shuffle
  update. The blue particles are eastbound
  the orange ones northbound. Between the lower-left and the
  upper-right the particles self-organize to form a diagonal pattern.
The white dashed lines delimit the transition zone
between the upper and lower triangular regions as discussed in
section\,\ref{subsection:tilt}.
Along the two entrance boundaries there are disordered boundary layers
of width $\xi\approx 50$.  
}  
\label{fig_alpha640_0.09}
\end{figure}

In simulations restricted
to linear sizes $M$ less than, say, a
few tens of lattice distances,
the particles seem to fill the intersection square largely randomly. 
However, as $M$ grows, one may discern more or less clear cut
local alignments of same-type particles along diagonals,
as shown in the snapshot in Fig.\,\ref{fig_alpha60_0.09}, where
$M=60$ and $\alpha=0.09$.
We therefore investigated what happens for much larger $M$.
It then appears that, far enough from the boundaries, the
particles form alternating stripes in the same way as observed for PBC.  
This is exemplified in Fig.\,\ref{fig_alpha640_0.09}, where $M=640$
and $\alpha=0.09$.
\vspace{2mm}

%%%%%%%%%%%%%%%%%%%%%%%%%%%%%%%%%%%
%%%%%%%%%%%%%%%%%%%%%%%%%%%%%%%%%%%
%%%%%%%%%%%%%%%%%%%%%%%%%%%%%%%%%%%
%%%%%%%%%%%%%%%%%%%%%%%%%%%%%%%%%%%

\begin{figure}
\begin{center}
\scalebox{.35}
{\includegraphics{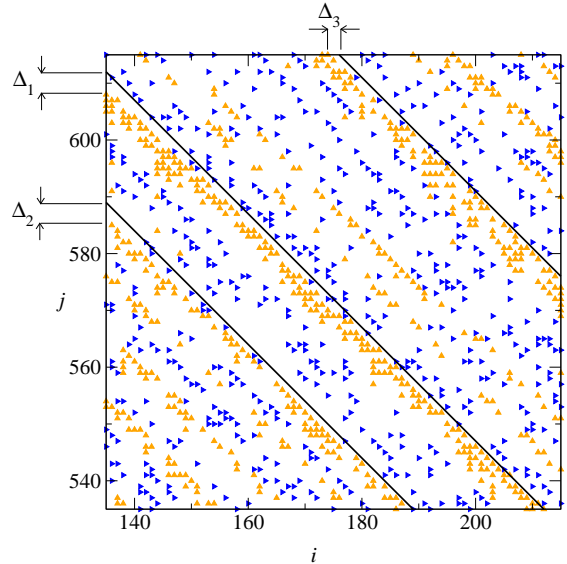}} 
\end{center}
\caption{\small Zoom on a region of Fig.\,\ref{fig_alpha640_0.09}
 showing the chevron pattern. 
 This snapshot has been taken for $M = 640$ and $\alpha = 0.09$. 
  The black solid lines are at an angle of $45^{\circ}$.
  The nonzero distances $\Delta_1,\Delta_2,$ and $\Delta_3$
  show that there is a negative $\Delta\theta$.
 Moreover, the northbound particles appear to be confined, statistically,
 to narrower strips than the eastbound ones.
} 
\label{fig_zoom640_0.09}
\end{figure}

Furthermore the following observations are of interest.
Along the two entrance boundaries
an $\alpha$ dependent {\it penetration depth\,} $\xi(\alpha)$
characterizes the distance that a randomly entering group of particles
needs to travel before it gets to self-organize into stripes. 
For $\alpha=0.09$ the figure shows that $\xi(\alpha)\approx 50$;
for $\alpha \to 0$ we found that this penetration depth diverges
roughly as $\xi(\alpha)\sim \alpha^{-1}$.
The organization into stripes reduces 
the probability of $\pE/\pN$ and $\pN/\pE$ blockings 
to below their values for a random 
distribution of particles, and it therefore increases the particles'
average velocity. 
Finally, we observe that the stripes are well-separated from
one another and move almost without any 
mutual penetration. This is borne out more clearly by the zoom
shown in Fig.\,\ref{fig_zoom640_0.09}, to which we will return later.

%%%%%%%%%%%%%%%%%%%%%%%%%%%%%%%%%%%
%%%%%%%%%%%%%%%%%%%%%%%%%%%%%%%%%%%
%%%%%%%%%%%%%%%%%%%%%%%%%%%%%%%%%%%
%%%%%%%%%%%%%%%%%%%%%%%%%%%%%%%%%%%

%%%%%%%%%%%%%%%%%%%%%%%%%%%%%%%%%%%%%%%%%%%%%%%%%%%%%%%%%%%%%%%%%%%%%%%%%%

\subsection{Stripe formation in the mean field model}
\label{section:mfanalysisobc}

{\it Nonlinear equations.\,\,}
We have numerically solved the nonlinear mean field equations (\ref{mfeqns2d})
on the intersection square, imposing time dependent
random densities as described in section \ref{secbcmf}
on the entrance sites along  the west and south boundaries,
and free exit conditions along the other two boundaries.
The system rapidly relaxes to
a stationary state independent of the
initial condition at time $t=0$; in our numerical resolution we
took for the latter
$\rhop_0(\brr)=\rhom_0(\brr)=\rhobar_0$ for all $\brr=(i,j)$ 
with $1\leq i,j\leq M$. 

The stripe structure becomes visible when we color 
a site blue or orange according to whether
$\rhop(\brr)$ is larger or smaller than $\rhom(\brr)$.
The result at time $t=1000$ is shown in Fig.\,\ref{fig_NLfield}.
Hence the mean field equations with OBC
lead to the same stripe formation as they did with PBC.
There is, similarly, a penetration depth along the entrance boundaries
within which the stripe structure has not yet developed.
The depth is here dependent on the average imposed boundary density
$\etabar$. 
\vspace{2mm}

{\it Linearized equations.\,\,}
The Green function of the linearized equations (\ref{mfeqns2dlin}) 
represents the response of the system
to an isolated boundary field $\etapm(\brr,t)$
acting at time $t=t_0$ on a single entrance site
$\brr=(0,j)$ or $\brr=(i,0)$. This Green function is easily
computed numerically, but may also be determined analytically.
The analytical calculation leads to an expression for several
quantities of interest,
among which the wavelength of the pattern of stripes,
\beq
\lambda = \sqrt{2} \pi/\arctan[(3-4\etabar)^{1/2}/(1-2\etabar)]\,.
\label{lambdaobc}
\eeq
The calculation is very lengthy and will be presented
elsewhere \cite{cividini_h2013}. 

%%%%%%%%%%%%%%%%%%%%%%%%%%%%%%%%%%%
%%%%%%%%%%%%%%%%%%%%%%%%%%%%%%%%%%%
%%%%%%%%%%%%%%%%%%%%%%%%%%%%%%%%%%%
%%%%%%%%%%%%%%%%%%%%%%%%%%%%%%%%%%%

{\it A remark.\,\,}
A comment is needed about the stability of Eqs.\,(\ref{mfeqns2d}).
The solution of these equations, after being perfectly reasonable, 
may develop a local
instability beginning by one of the densities $\rhop_t$ or
$\rhom_t$ becoming negative at a particular site $\brr$,
after which in a few time steps the solution tends to infinity in that
region. 
For densities around $0.100$, which are among the highest that we
consider here, and for $M=500$, this typically happens
after 5000 to 15\,000 time steps. For lower densities it is rarer.
The average $\langle\ldots\rangle$ at time $t$ is therefore 
defined as restricted to all 
histories for which no instability has occurred up to that time.
The instability is suppressed if we add higher order nonlinear terms
to Eqs.\,(\ref{mfeqns2d}), {\it e.g.} by 
replacing $\rhop_t$ with $1-\ee^{-\rhop_t}$
in the first of Eqs.(\ref{mfeqns2d}) and similarly  $\rhom_t$ with $1-\ee^{-\rhom_t}$
in the second one. In that case one easily shows that for initially
positive densities and appropriate boundary conditions 
the solution must stay positive at all times.
Simulations with this exponential density dependence
give results very close to the ones obtained 
from Eqs.\,(\ref{mfeqns2d}).

%%%%%%%%%%%%%%%%%%%%%%%%%%%%%%%%%%%%%%%%%%%%%%%%%%%%%%%%%%%%%%%%%%%%%%%%%%%%%%

\subsection{Chevron effect}
\label{subsection:tilt}

Having found that both the particle and the mean field model
are subject to the pattern formation instability,
we continue in this subsection our investigation of the
pattern structure. 

\begin{figure}
\begin{center}
\scalebox{.32}
{\includegraphics{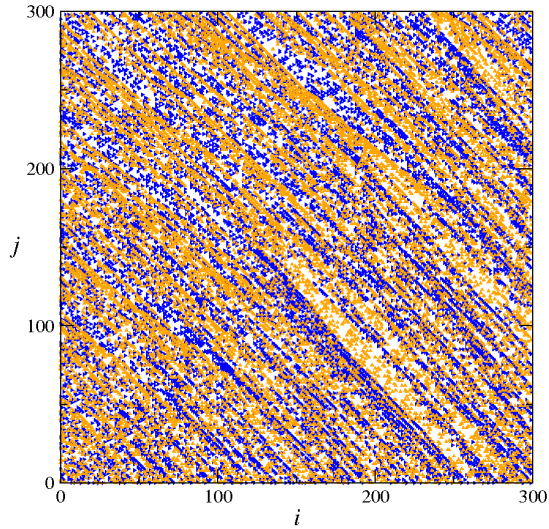}}
\end{center}
\caption{\small Snapshot of the stationary state of the intersection
  square of a particle system
  with $M = 300$ and $\alpha = 0.15$ and subject to alternating parallel
  update. The color code is the same as in Fig.\,\ref{fig_alpha640_0.09}.
The disordered boundary layers along the entrance boundaries are much
narrower here than in Fig.\,\ref{fig_alpha640_0.09}.
}  
\label{fig_AP300_0.15}
\end{figure}

Closer examination of Fig.\,\ref{fig_alpha640_0.09} (particle model
with frozen shuffle update) and
Fig.\,\ref{fig_NLfield} (mean field model with
density boundary conditions)  
reveals an effect just barely visible to the eye
(it is better visible if the paper is held flat!), 
namely that the angle $\theta$ of the striped pattern,
is {\it not exactly equal}
{\it to\,} $45^{\circ}$ but differs systematically from it by an 
amount $\Delta\theta(\brr)$ which is of the order of a degree. 
This angle difference $\Delta\theta(\brr)$ is negative above the axis 
of symmetry and positive below it, so that the stripes acquire the
character of chevrons, 
as schematically represented in Fig.\,\ref{fig_stripesbc}b;  
we will therefore call this the `chevron effect' and refer to 
$\Delta\theta(\brr)$ as the `chevron angle'. 
We have verified that the chevron effect also occurs in
the particle model with alternating parallel update and show proof of
this in Fig.\,\ref{fig_AP300_0.15}.

From the fact that it appears
under all these different conditions we conclude that it is a robust
property of a wide class of intersecting flow models.
We note, however, that no sign of the chevron effect 
appeared in the linear stability analysis mentioned in section 
\ref{secinstability2} above; hence the effect is essentially connected
to the nonlinearity of the evolution equations.

We will now investigate the quantitative relation between
$\theta(\brr)$ and the control parameters, $\alpha$ or $\etabar$.
A prerequisite is that we define this angle
in a way allowing us to quantify it
operationally in the simulations.
We will successively discuss two different algorithms that we
conceived for this purpose, and that we termed the 
{\it crest method\,} and the {\it velocity ratio\,} method.

%%%%%%%%%%%%%%%%%%%%%%%%%%%%%%%%%%%%%%%%%%%%%%%%%%%%%%%%%%%%%%%%%%%%%%%%%%%%%%

\subsubsection{Crest  method}
\label{secchevronangle}

%%%%%%%%%%%%%%%%%%%%%%%%%%%%%%%%%%%
%%%%%%%%%%%%%%%%%%%%%%%%%%%%%%%%%%%
\begin{figure}
\begin{center}
\scalebox{.4}
{\includegraphics{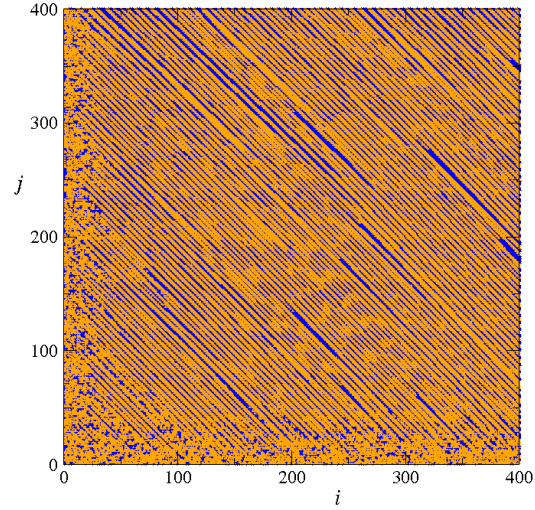}} 
\end{center}
\caption{\small 
Snapshot representing a typical stationary state configuration of
the density fields in an intersection square of
size $M=400$ with OBC, 
subject to a fluctuating boundary density of average
$\etabar=0.06$. 
A site $\brr$ is colored blue or orange according to whether
$\rhop(\brr)$ is larger or smaller than $\rhom(\brr)$. 
}
\label{fig_NLfield}
\end{figure}
%%%%%%%%%%%%%%%%%%%%%%%%%%%%%%%%%%%
%%%%%%%%%%%%%%%%%%%%%%%%%%%%%%%%%%%

%%%%%%%%%%%%%%%%%%%%%%%%%%%%%%%%%%%
%%%%%%%%%%%%%%%%%%%%%%%%%%%%%%%%%%%
%%%%%%%%%%%%%%%%%%%%%%%%%%%%%%%%%%%
%%%%%%%%%%%%%%%%%%%%%%%%%%%%%%%%%%%

%%%%%%%%%%%%%%%%%%%%%%%%%%%%%%%%%%%
%%%%%%%%%%%%%%%%%%%%%%%%%%%%%%%%%%%
\begin{figure}
\begin{center}
\scalebox{.4}
{\includegraphics{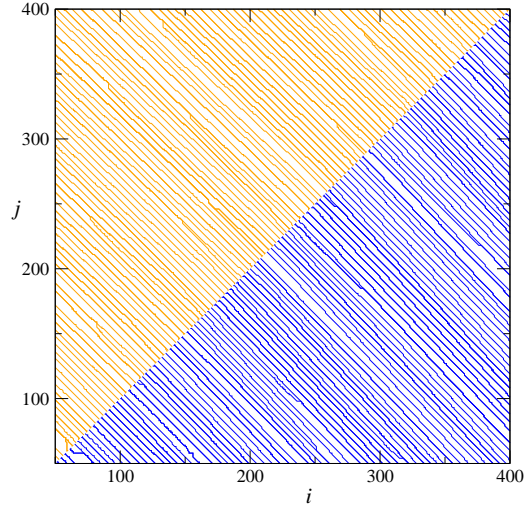}} 
\end{center}
\caption{\small 
Result of the crest construction algorithm described in section
\ref{secchevronangle},
applied to the density field of Fig.\,\ref{fig_NLfield}.
For the value $\etabar=0.06$ that was used here,
the slope of the blue crests (lower triangle) and the orange crests
(upper triangle) differs by about $2\Delta\theta_0=2.4^{\circ}$.
Layers of width 50 along both entrance boundaries were left out of the
construction algorithm. 
} 
\label{fig_NLcrests}
\end{figure}
%%%%%%%%%%%%%%%%%%%%%%%%%%%%%%%%%%%
%%%%%%%%%%%%%%%%%%%%%%%%%%%%%%%%%%%

%%%%%%%%%%%%%%%%%%%%%%%%%%%%%%%%%%%
%%%%%%%%%%%%%%%%%%%%%%%%%%%%%%%%%%%
\begin{figure}
\begin{center}
\scalebox{.36}
{\includegraphics{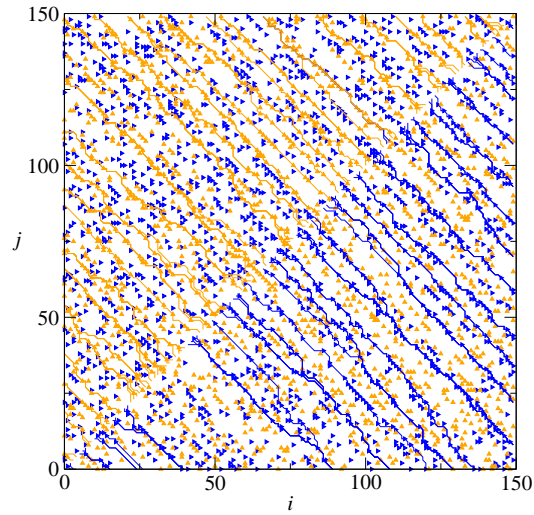}} 
\end{center}
\caption{\small Snapshot of the intersection square for a particle system
with $\alpha = 0.09$ and $M = 150$, obtained with frozen shuffle update.
Superimposed are the crests constructed
by means of the algorithm of section \ref{secchevronangle}.
} 
\label{fig_partcrests}
\end{figure}
%%%%%%%%%%%%%%%%%%%%%%%%%%%%%%%%%%%
%%%%%%%%%%%%%%%%%%%%%%%%%%%%%%%%%%%

In order to assign a numerical value to the slope
$\theta$, we developed an algorithm that closely imitates
what visual inspection does: given a configuration
of the particle occupation numbers, $\{\np(\brr),\nm(\brr)\}$, or 
of the density fields,  $\{\rhop(\brr),\rhom(\brr)\}$,
it follows the clearly visible crests over a certain distance. 
This algorithm is easiest to apply to the 
density fields, their variables being continuous.
It is in that case composed of the following steps.

Each diagonal site $\brr=(k,k)$ occupied by an eastbound particle
is taken as the initial site of a crest
to be constructed stepwise towards the south-east.
If the current crest end is at $(i,j)$,
then the next site on the crest will be one of the three sites
$(i,j-1)$, $(i+1,j-1)$, or $(i+1,j)$, 
whichever has the largest value of $\rhop$. 
The construction ends when the crest reaches the south or east
boundary of the intersection square; it may also be restricted to a
smaller square resulting from the exclusion of 
the boundary layers. 
The end-to-end distance of a crest with initial site
$(k,k)$ is a vector that we will denote $\bc(k)$. Let
$\bC^{\pE} = \sum_k^{\pE}\bc(k)$, where the superscript on the 
summation sign denotes restriction to initial sites occupied by an 
eastbound particle.
Finally, $\theta$ is taken to be the angle of $\bC^{\pE}$.
It represents an average over the {\it lower\,} triangular half of the
intersection square. The precision may be increased by repeating the
measurement and adding the $\bC^{\pE}$ obtained from a sequence of
configurations. 

Symmetrically, starting from the diagonal sites occupied by a
northbound particle, a similarly
constructed vector $\bC^{\pN}$ leads to a value of $\theta$ averaged
over the {\it upper\,} triangular half of the intersection square.
Fig.\,\ref{fig_NLcrests} shows the crests constructed this way
for the mean field configuration of Fig.\,\ref{fig_NLfield}.

In a particle configuration the local densities $n^{\pEN}(\brr)$ are binary
variables equal to 0 or 1.
Before applying the crest algorithm to it,
we first create the two density fields $\rhop$ and $\rhom$ such that
$\rhopm(\brr) = n^{\pEN}(\brr)$.
In order to lift any degeneracies
we then apply three diffusion steps in each of which,
for the two fields separately,
each site distributes a fraction $\epsilon$ of its 
density content equally over its four neighboring sites
(in practice we took $\epsilon = 0.1$).
After that, the above algorithm can be applied.
Fig.\,\ref{fig_partcrests} shows simultaneously
a snapshot of a particle system having $M=150$ and $\alpha=0.09$, and
the crests constructed from it.

The same algorithm may serve to construct 
crests locally and to determine, through an ensemble average, 
the local values $\theta(\brr)$.
Clearly, too, this algorithm, although natural and making sense, is
not unique and different but equally reasonable algorithms
might well lead to slightly modified values of the angles $\theta(\brr)$. 

%%%%%%%%%%%%%%%%%%%%%%%%%%%%%%%%%%%%%%%%%%%%%%%%%%%%%%%%%%%%%%%%%%%%%%%%%%%%

\subsubsection{Velocity ratio method}
\label{secvelocityratio}

The velocity ratio method for determining the angle
$\theta(\brr)$ is based on an elementary theoretical 
consideration. 
Let $\vp(\brr)$ and $\vm(\brr)$ be the average 
eastward and northward velocities, respectively, on site $\brr$
in the stationary state. The velocities refer to particles or to
fields, as the case may be.
Under the sole hypothesis, borne out rather well 
both in the simulations of the particle model and in the numerical
solution of the mean field model,
that the stripes are mutually impenetrable, 
the existence of such moving stripes is possible only if locally
their angle of inclination $\theta(\brr)$ is related to their
average velocities $\vpm(\brr)$ by
\beq
\tan\theta(\brr)=\frac{\vm(\brr)}{\vp(\brr)}\,.
\label{xtanthetavv}
\eeq

In the particle model the velocities are defined by 
\beq
\vpm(\brr) = \frac{\Jpm(\brr)}{\langle \npm(\brr) \rangle},
\label{defv}
\eeq
where $\Jpm(\brr)$ is the stationary current of $\pm$ particles on site $\brr$.
For the mean field model similar equations hold with the $n$'s
replaced with $\rho$'s. 

Things simplify in the special case
\footnote{The particle model with OBC
is the prime example.}
where the boundary conditions impose the same stationary state 
current $J$ in each horizontal and vertical lane, 
and hence on each site. In that case, combining (\ref{xtanthetavv}) and (\ref{defv}) 
yields, in lieu of (\ref{xtanthetavv}),
an expression for the local slope $\theta(\brr)$
solely in terms of the two local densities, 
\beq
\tan\theta(\brr)
= \frac{\langle\np(\brr)\rangle}{\langle\nm(\brr)\rangle}\,.
\label{xtantheta}
\eeq
Setting $\theta = \frac{\pi}{4}+\Delta\theta$
and expanding (\ref{xtantheta}) to linear order in $\Delta\theta(\brr)$ yields
\beq
\Delta\theta(\brr) \simeq 
\frac{\la\np(\brr)\ra-\la\nm(\brr)\ra}{2\la\nm(\brr)\ra}\,,
\label{xDthetaJ}
\eeq
valid only if $\Jpm(\brr) = J$ is uniform.
Eqs.\,(\ref{xtantheta}) and (\ref{xDthetaJ}) show, 
in particular, that there can be a nonzero chevron angle
$\Delta\theta(\brr)$ only if the local densities of the two
species are different.
Our simulations of the particle model with OBC
indeed show this density difference.

%%%%%%%%%%%%%%%%%%%%%%%%%%%%%%%%%%%%%%%%%%%%%%%%%%%%%%%%%%%%%%%%%%%%%%%%%%%%

\subsubsection{Comparison}
\label{seccomparison}

We will take Eq.\,(\ref{xtanthetavv}) as the definition of $\theta(\brr)$.
However, it should be remembered
that when the impenetrability hypothesis is violated,
the quantity $\theta(\brr)$ defined by (\ref{xtanthetavv})
loses its interpretation as the slope of a stripe.
This happens near the two entrance boundaries:
the disorderly structure within a distance $\xi$ from these boundaries
renders the slope ill-defined, even though
blind application of Eq.\,(\ref{xtanthetavv}) gives a precise value.

\begin{figure}
\begin{center}
\scalebox{.35}
{\includegraphics{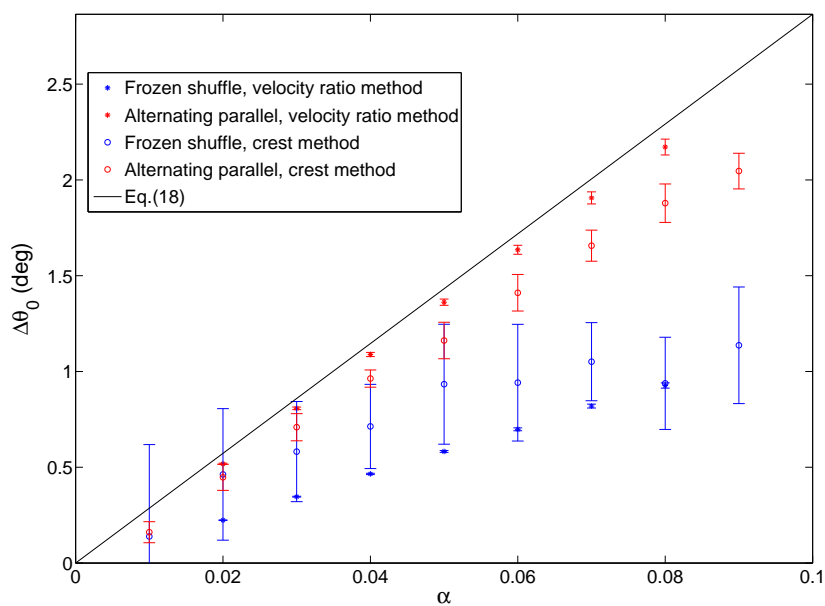}} 
\end{center}
\caption{\small 
Chevron angle
$\Delta\theta_0$ as a function of $\alpha$ for open boundary conditions, 
obtained for the two update schemes and
by the two measuring methods described in section \ref{subsection:tilt}
and the two update schemes discussed in section \ref{secupdateschemes},
obtained in a system of linear size $M=640$.
The solid black line corresponds 
to the theoretical expression (\ref{approxtanthetalin})  
derived for alternating parallel update on the basis of the special
class of stripes of Fig.\,\ref{fig_Dtheta}.
} 
\label{fig_dth0}
\end{figure}

When well-defined stripes do exist, one expects
the crest algorithm and the velocity ratio method to yield,
if not identical, then at least closely similar results.
For all situations that we have considered this turns out to be the
case.
A comparison of the two determinations of $\theta(\brr)$ will be 
made in section \ref{section:angleDtheta} (see Fig.\,\ref{fig_dth0}) 
and in section \ref{section:cbc} (see Fig.\,\ref{fig_cylDthetafin}).

%%%%%%%%%%%%%%%%%%%%%%%%%%%%%%%%%%%%%%%%%%%%%%%%%%%%%%%%%%%%%%%%%%%%%%%%%%%%%

\subsubsection{Space dependence of  $\Delta\theta(\brr)$}
\label{secspacedep}

%%%%%%%%%%%%%%%%%%%%%%%%%%%%%%%%%%%%
%%%%%%%%%%%%%%%%%%%%%%%%%%%%%%%%%%%%
\begin{figure}
\begin{center}
\scalebox{.30}
{\includegraphics{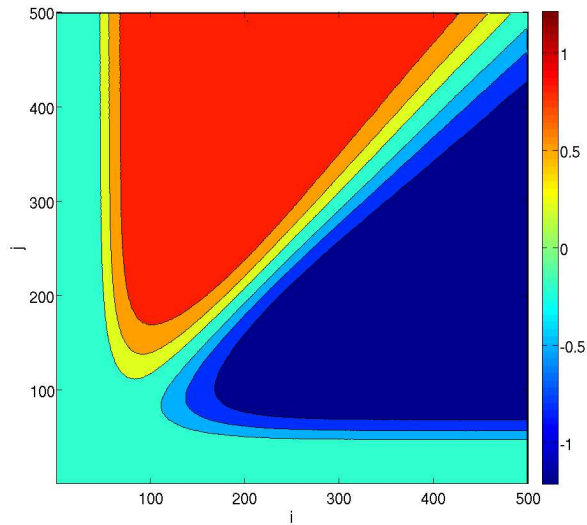}}
\end{center}
\caption{\small Level plot of the space dependent chevron angle 
  $\Delta\theta(i,j)$ (in degrees), in the stationary state,
  obtained from the mean field equations (\ref{mfeqns2d})
  by the velocity ratio method, on a
  square lattice of linear size $M=500$ with open boundary conditions and
  with $\etabarp=\etabarm=0.06$.
  } 
\label{fig_Dthetalevel}
\end{figure}
%%%%%%%%%%%%%%%%%%%%%%%%%%%%%%%%%%%%
%%%%%%%%%%%%%%%%%%%%%%%%%%%%%%%%%%%%

Fig.\,\ref{fig_Dthetalevel} shows  
the space dependence
of the chevron angle in the intersection square with OBC
after the system has attained its stationary state.
It has been obtained by the velocity ratio method, Eq.\,(\ref{xtanthetavv}),
from the solution of the nonlinear mean field equations (\ref{mfeqns2d}). 
An average over $10^5$ time steps has been performed. 
This figure provides quantitative support for
the observation, already strongly suggested by Fig.\,\ref{fig_alpha640_0.09},
that we may divide the intersection square
into reasonably well-defined regions
according to the value of $\Delta\theta(\brr)$.

(a) In layers of width $\xi$
along the two entrance boundaries 
$\Delta\theta(\brr)$ fluctuates around zero. One of the 
particle types is here clearly not organized into stripes. 
As a consequence, the impenetrability hypothesis is violated and
the values obtained for $\Delta\theta(\brr)$ 
in these boundary layers,
although unambiguously defined by (\ref{xtanthetavv}),
do not represent angles of inclination.

(b) There are, clearly visible in Fig.\,\ref{fig_Dthetalevel},
two triangular regions where
$\Delta\theta (\brr)$ is close to constant. 
We will denote the value of this constant by $\pm \Delta\theta_0$
in the lower and upper triangle, respectively.

(c) Along the axis of symmetry there is a transition zone where 
$\Delta\theta(\brr)$ passes from $-\Delta\theta_0$ 
above the axis to $+\Delta\theta_0$ below it. 
In Fig.\,\ref{fig_alpha640_0.09}
this variation causes a slight rounding at the tips of the chevrons.
The transition zone is clearly visible in Fig.\,\ref{fig_Dthetalevel}
and has been indicated by heavy white dashed lines
in Fig.\,\ref{fig_alpha640_0.09}.

\begin{figure}
\begin{center}
\scalebox{.30}
{\includegraphics{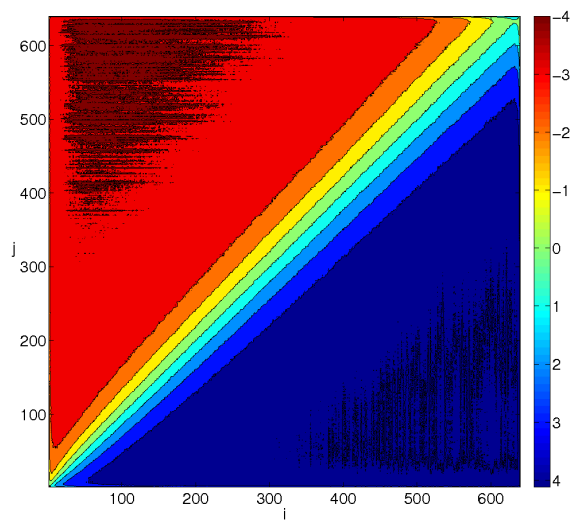}}
\end{center}
\caption{\small Level plot of the space dependent chevron angle 
  $\Delta\theta(i,j)$ (in degrees), in the stationary state,
  obtained by the velocity ratio method for a particle model with 
  alternating parallel update on a
  square lattice of linear size $M=640$ with open boundary conditions and
  with $\alpha=0.15$.
  } 
\label{fig_Dthetalevelpart}
\end{figure}

In Fig.\,\ref{fig_Dthetalevelpart} we show the analogous plot
for a particle model on a square lattice of linear size $M=640$.
Regions similar to those in Fig.\,\ref{fig_alpha640_0.09} are visible.
For this model we have chosen $\alpha=0.15$, 
a value just below its jamming point, 
which gives rise to a chevron angle $\Delta\theta_0\approx 4^\circ$.

%%%%%%%%%%%%%%%%%%%%%%%%%%%%%%%%%%%%
%%%%%%%%%%%%%%%%%%%%%%%%%%%%%%%%%%%%
%%%%%%%%%%%%%%%%%%%%%%%%%%%%%%%%%%%%
%%%%%%%%%%%%%%%%%%%%%%%%%%%%%%%%%%%%

%%%%%%%%%%%%%%%%%%%%%%%%%%%%%%%%%%%
%%%%%%%%%%%%%%%%%%%%%%%%%%%%%%%%%%%
%%%%%%%%%%%%%%%%%%%%%%%%%%%%%%%%%%%
%%%%%%%%%%%%%%%%%%%%%%%%%%%%%%%%%%%

%%%%%%%%%%%%%%%%%%%%%%%%%%%%%%%%%%%%%%%%%%%%%%%%%%%%%%%%%%%%%%%%%%%%%%%%%%%

\subsubsection{The angle $\Delta\theta_0$}
\label{section:angleDtheta}

For a sequence of values of $\alpha$ 
we simulated a particle model having $M = 640$, 
employing successively the two different update schemes 
discussed in section \ref{secupdateschemes},
and determined $\theta(\brr)$
by the two different methods discussed above.
Fig.\,\ref{fig_Dthetalevel} shows the values
for $\Delta\theta(\brr)$ obtained by the velocity ratio method.
In each simulation there appear to be two triangular
regions, symmetric about the diagonal and roughly coinciding with the
red and dark blue regions in figure \ref{fig_Dthetalevelpart},
where $\Delta\theta(\brr)$ is close to what appears like a limit 
value. We called the result of the spatial averages over these 
regions $\pm\Delta\theta_0$. 
The crest method was applied to the same regions and its outcome for
a single particle configuration was
averaged over repeated determinations at different instants of time.

The results of both methods are plotted in Fig.\,\ref{fig_dth0}. 
The error bars 
represent the statistical standard deviation of
the averaged values.
It appears that in all cases there is a chevron effect and that 
its magnitude $\Delta\theta_0$
depends linearly on the control parameter $\alpha$.

In addition, numerical solution of the mean field model, not shown here,
brings out a similar linear dependence of $\Delta\theta_0$ on the
control parameter $\etabar$.
Specifically, we established that
\beq 
\Delta\theta_0 \simeq c\alpha, \qquad 
\Delta\theta_0\simeq c^{\prime}\etabar
\label{xDthetalin} 
\eeq
with $c \approx 12^\circ$ 
for the frozen shuffle update and $c \approx 26^\circ$ for the
alternating parallel update in the regime $\alpha\lesssim 0.10$; 
while $c^{\prime} \approx 21^{\circ}$
for the mean field model in the same regime, $\etabar\lesssim 0.10$.

%%%%%%%%%%%%%%%%%%%%%%%%%%%%%%%%%%%%%%%%%%%%%%%%%%%%%%%%%%%%%%%%%%%%%%%%%%%%%

\subsubsection{Definition of $\Delta\theta_0$ and limit $M \rightarrow \infty$}
\label{section:Minfty}

\begin{figure}
\begin{center}
\scalebox{.4}
{\includegraphics{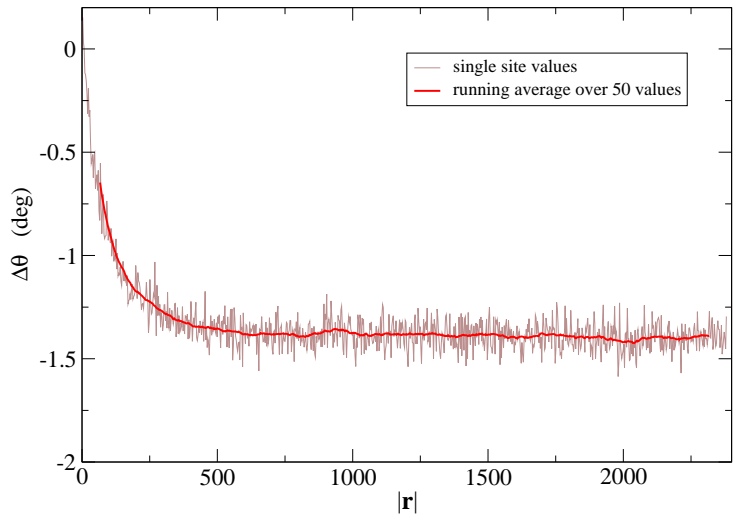}} 
\end{center}
\caption{\small 
Chevron angle $\Delta\theta(\brr)$ 
determined by the velocity ratio method
as a function of $|\brr|=(i^2+j^2)^{1/2}$
along the the line $j/i=\tan(3\pi/8)$, 
taken on one site $j$ in each column $i$, 
for $M=2200$ and $\alpha=0.05$, in a particle system 
with open boundary conditions 
subject to alternating parallel update. 
The light (brown) curve is the average over $2\times 10^6$ time steps
after 5000 time steps have been discarded.
The dark (red) curve is a running average over 50 points.
}
\label{fig_minf}
\end{figure}

It is natural to ask what the stationary state will look like
in the limit $M \rightarrow \infty$. 
However, for nonequilibrium systems like this one
there appears to be little, if any, theoretical 
guidance to answer this question\footnote
{A similar question
was briefly discussed in Ref.\,\cite{hilhorst_a2012}.}.
We therefore performed particle simulations 
for system of very large linear sizes, up to $M = 2900$, 
employing the alternating parallel update scheme, 
with the purpose of studying the behavior of the chevron angle
$\Delta\theta(\brr)$ at large distances $|\brr|$.
Fig.\,\ref{fig_minf} shows this angle 
for a system having $M=2200$ and for $\alpha=0.05$
along the line
$j/i=\tan(3\pi/8)$, which bisects the upper triangular region.

The figure shows that along this line, $\Delta\theta(\brr)$ exhibits
first of all a steep initial decrease with the distance from the origin.
Once the penetration depth is reached, it remains
confined to a narrow range around $-1.39^{\circ}$. 
The fluctuations around this plateau value
are compatible with $\Delta\theta(\brr)$ tending to a constant
  along this line; however, we have no theoretical argument to exclude
a very slow decay towards zero.
We will therefore abstain from what one might have liked to do, namely
defining $\Delta\theta_0$ 
as the $|\brr|\to\infty$ limit of $|\Delta\theta(\brr)|$ 
in an appropriate direction.
Instead, we will satisfy ourselves with the procedure
of the preceding subsections, which amounts to
identifying $\Delta\theta_0$ with the plateau value first reached 
when $|\brr|$ exits the boundary layer.

%%%%%%%%%%%%%%%%%%%%%%%%%%%%%%%%%%%
%%%%%%%%%%%%%%%%%%%%%%%%%%%%%%%%%%%
%%%%%%%%%%%%%%%%%%%%%%%%%%%%%%%%%%%
%%%%%%%%%%%%%%%%%%%%%%%%%%%%%%%%%%%

%%%%%%%%%%%%%%%%%%%%%%%%%%%%%%%%%%%%%%%%%%%%%%%%%%%%%%%%%%%%%%%%%%%%%%%%%%%%%

\section{Chevron effect: theoretical arguments}
\label{section:chevrontheory}

%%%%%%%%%%%%%%%%%%%%%%%%%%%%%%%%%%%
%%%%%%%%%%%%%%%%%%%%%%%%%%%%%%%%%%%
\begin{figure}
\begin{center}
\scalebox{.30}
{\includegraphics{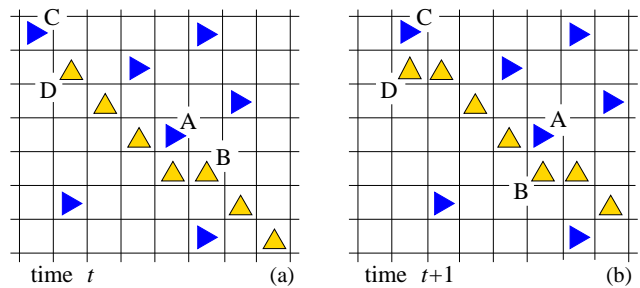}}
\end{center}
\caption{\small Mechanism causing the deviation $\Delta\theta$
of a stripe of northbound particles (\orangew{$\blacktriangle$}) 
in the upper triangular region. The update is alternating parallel;
during the time step shown first all eastbound and then all 
northbound particles move, provided their target site is empty.
}
\label{fig_Dtheta}
\end{figure}
%%%%%%%%%%%%%%%%%%%%%%%%%%%%%%%%%%%
%%%%%%%%%%%%%%%%%%%%%%%%%%%%%%%%%%%

We refer to the zoom, shown in Fig.\,\ref{fig_zoom640_0.09},  
on an area located in the upper triangular region 
of Fig.\,\ref{fig_alpha640_0.09}.
The zoom makes clear that in this region
there is an important asymmetry in the
spatial distribution of the eastbound and the northbound particles:
the stripes of the former are dense and narrow, 
whereas those of the latter are sparse and wide. 
A consequence, visible even if barely so, is that
the upper triangular region in Fig.\,\ref{fig_alpha640_0.09}
looks bluish and the lower triangle more orange-like.
The asymmetry observed here
offers the clue to an elementary theory of the chevron effect.

The core of the problem is to show that the system 
is capable of sustaining modes of propagation 
in which the stripes have a slope different from $45^\circ$.
Let us consider what happens near
the entrance boundary of the eastbound particles, that is, 
for $i\approx\xi$ but $j\gg\xi$. 
Near this boundary the eastbound particles (\bluew{$\blacktriangleright$}),
after having entered the intersection square randomly,  
fill the space offered to them 
between the northbound stripes (\orangew{$\blacktriangle$})
also largely randomly.
This suggests to consider the special 
class of northbound stripes exemplified
in Fig.\,\ref{fig_Dtheta}a. The stripes 
consist of straight segments at an angle of $45^\circ$,
concatenated by `kinks' such as the one that occurs in
Fig.\,\ref{fig_Dtheta}a at the level of particle B,
and that is associated with the presence of the 
eastbound particle A.
The other eastbound particles in Fig.\,\ref{fig_Dtheta}a 
occupy random positions.
Now, one time step of alternating parallel update applied  
to the configuration of Fig.\,\ref{fig_Dtheta}a, will take it to
that of Fig.\,\ref{fig_Dtheta}b. This may be seen in detail as follows.
We attempt to move in parallel
first all eastbound and then all northbound particles.
We see that during the time step from $t$ to $t+1$ none of the 
eastbound particles is blocked. 
In particular, the moves of A and C 
block B and D, respectively.
Consequently, after the unblocked northbound particles have also
moved, the kink associated with A has been displaced one lattice
distance to the right along the northbound stripe, 
and C has created a new kink at the beginning of that
stripe, at the level of particle D; this is represented 
in Fig.\,\ref{fig_Dtheta}b.
In subsequent time steps 
particles A and C will both travel from left to right along the stripe,
each of them taking its associated kink along,
and the connected structure of the stripe will be preserved.

If the set of kinks has a linear density $\rho_{\rm kink}$
along the stripe, the average stripe angle $\theta$ will be
given by $\tan\theta = 1-\rho_{\rm kink}$. 
Since $\rho_{\rm kink}$ also represents the fraction of blocked moves
of the northbound particles, the stripe's
speed will be $\vm=1-\rho_{\rm kink}$.
By contrast, the eastbound particles move at speed $\vp=1$.

In the example of Fig.\,\ref{fig_Dtheta} we note that
a uniformly random spatial distribution of the particles 
with density $\rhop$ would lead to 
$\rho_{\rm kink}=\rhop$. Because $\vp=1$ we have, moreover that 
$\rhop=J$.
Using the above expressions, valid in the special situation of
Fig.\,\ref{fig_Dtheta}, in Eq.\,(\ref{xtanthetavv}) 
we are led to a fully explicit expression for the angle, namely
\beq
\tan\theta(\brr) = 1-J,
\label{approxtantheta}
\eeq
in which $J(\alpha)$ is given by Eq.\,(\ref{xJ}), and which results in
an angle $\theta$ that is independent of $\brr$
within the region where the preceding approximations apply. 
To lowest order in $\Delta\theta(\brr)$ and $\alpha$ this yields,
converted to degrees,
\beq
\Delta\theta(\brr) = \frac{\alpha}{2} \left( \frac{180}{\pi} \right)^{\circ},
\label{approxtanthetalin}
\eeq
which has been plotted in Fig.\,\ref{fig_dth0}
as a comparison with the plateau values $\Delta\theta_0$. 
Since a correlated distribution of the eastbound 
particles would lead to a lower
$\rho_{\rm kink}$, we expect that Eq.\,(\ref{approxtantheta}), while giving
the correct order of magnitude, overestimates the slope; this
is confirmed by the figure.

Hence the special class of stripes
depicted in Fig.\,\ref{fig_Dtheta} demonstrates  
the most distinctive ingredient of the chevron effect:
the existence of a nonlinear mode consisting of a stripe
with an average slope different from $45^\circ$
and two distinct speeds of propagation, $\vm<\vp$.
We must expect similar modes to be present for
a wide class of models, including
the original particle model with frozen shuffle update
as well as the mean field model;
for these models an explicit analysis would however be
much more difficult.

In the case of the particle model we show in a complementary study
\cite{cividini_a2013}
how an eastbound particle may get localized in the wake of another
one when the two are immersed in a sea of northbound particles;
the wake having the same slope as the striped mode described here.
\vspace{2mm}

The rule that we may derive from these considerations is the
following.

\begin{Rule}
\label{ruleangle}
At the interface where a disordered species A of density $\rho_A$
penetrates into a perpendicular traveling and
diagonally ordered species B,
the speed at which the B diagonals advance
is reduced from $1$ to $1-\rho_A$.
Moreover, 
the (acute) angle between the diagonals 
and the direction of propagation of the disordered species 
is reduced from $\frac{\pi}{4}$ by an amount $\frac{1}{2}\rho_A$
(which corresponds to $\Delta\theta_0=\pm\frac{1}{2}\rho_A$, as the
case may be).
\end{Rule}

%%%%%%%%%%%%%%%%%%%%%%%%%%%%%%%%%%%%%%%%%%%%%%%%%%%%%%%%%%%%%%%%%%%%%%%%%%%%%

\section{Chevron effect on a cylinder}
\label{section:cbc}

As shown above, the chevron effect is present for open boundary
conditions (OBC) but not for periodic ones (PBC).
We will now study it in the intermediate case of cylindrical boundary conditions
(CBC, open for the eastbound and periodic for the northbound particles) 
and show that by controlling the 
asymmetry between the two directions we will 
better our understanding of the chevron effect.

An advantage  of the cylindrical geometry 
is the translational invariance,
in the statistical sense, in the vertical direction. 
As a consequence, by averaging the quantities of interest 
over the vertical coordinate, we will obtain higher precision results
than for the two-way open system.

We will present this cylinder study only for the mean field model 
(\ref{mfeqns2d}) and briefly comment in the end on
analogous results for the particle model.
For CBC the density of the northbound particles
is strictly conserved in each column separately.
We will set its average equal to $\rhobarm$.
The initial values $\rhom_0(\brr)$ were drawn as i.i.d. 
variables $\rho$ from
the distribution $p(\rho;\rhobarm)$ of Eq.\,(\ref{dprho}).

\begin{figure}
\begin{center}
\scalebox{.35}
{\includegraphics{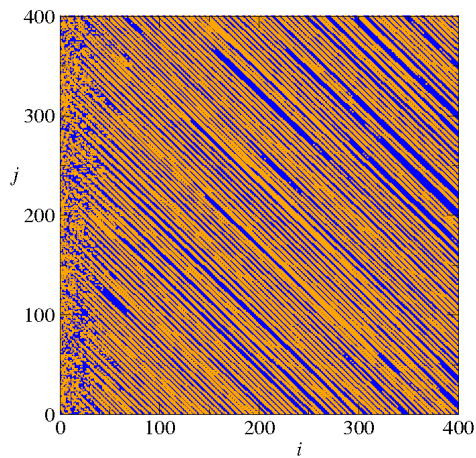}}
\end{center}
\caption{\small Snapshot representing the density fields in the
  intersection square of size $M=400$ and subject to cylindrical
  boundary conditions with control parameters $\rhobar^\pN=0.050$ and
  $\etabar^\pE=0.055$. The color code is as in Fig.\,\ref{fig_NLfield}.
  A disordered boundary layer along the west entrance is clearly
  visible. The striped pattern in the bulk has an average slope
  of $45^\circ - \Delta\theta_0$; for the pattern
  shown the crest method gives $\Delta\theta_0\approx 1.13^\circ$.
  } 
\label{fig_NLcylfield}
\end{figure}

For the eastbound particles the control parameter is 
the average density $\etabarp$ of the boundary noise. 
This study therefore has the independent parameters
$\etabarp$ and $\rhobarm$.
Fig.\,\ref{fig_NLcylfield} represents a snapshot of
the density fields in the stationary state
for the particular set of values $\rhobar^\pN=0.050$ and
$\etabar^\pE=0.055$ on an interaction square of linear dimension
$M=400$.   The configuration was obtained by numerical solution of the
nonlinear mean field equations (\ref{mfeqns2d}) with CBC during $2000$
time steps; the memory of the initial state has then disappeared
and the system has entered a stationary state.

%%%%%%%%%%%%%%%%%%%%%%%%%%%%%%%%%%%%
%%%%%%%%%%%%%%%%%%%%%%%%%%%%%%%%%%%%
%%%%%%%%%%%%%%%%%%%%%%%%%%%%%%%%%%%%
%%%%%%%%%%%%%%%%%%%%%%%%%%%%%%%%%%%%

%%%%%%%%%%%%%%%%%%%%%%%%%%%%%%%%%%%%%%%%%%%%%%%%%%%%%%%%%%%%%%%%%%%%%%%%%%

\subsection{Chevron effect  in the stationary state}
\label{secchevronstationary}

Although not easily visible to the eye, the stripes in 
Fig.\,\ref{fig_NLcylfield}
are at an angle less than $45^\circ$, in the way 
schematically shown in Fig.\,\ref{fig_stripesbc}c.
Along the west entrance there is a disordered boundary zone.
We will now ask about the chevron angle $\Delta\theta$
as a function of the column index $i$.

\begin{figure}
\begin{center}
\scalebox{.35}
{\includegraphics{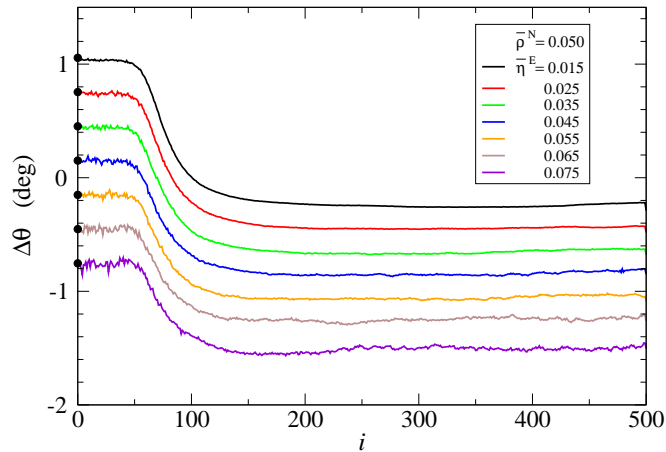}}
\end{center}
\caption{\small Chevron angle $\Delta\theta$ as a function of the
  column index $i$,
  obtained by simulation of the mean field equations (\ref{mfeqns2d})
  and measured by means of the velocity ratio method, on a
  square lattice of linear size $M=500$ with cylindrical boundary conditions.
  The northbound particles have a fixed
  density $\rhobar^\pN=0.050$. Curves are shown for various boundary
  densities $\etabar^\pE$ of the eastbound particles.
  The dots for $i=1$ are the boundary values predicted by 
  Eq.\,(\ref{xDeltathetacyl1}).
  } 
\label{fig_cylDthetacolumn}
\end{figure}

Fig.\,\ref{fig_cylDthetacolumn} shows the stationary state values of\,%
\footnote{We write $X(i)$ instead of $X(\brr)=X(i,j)$ for any quantity
  $X$ depending only on the column index $i$.}
$\Delta\theta(i)$, obtained by the velocity ratio method of section
\ref{secvelocityratio}, that is, from Eq.\,(\ref{xtanthetavv}).
Every curve was obtained as an average over at least 60 field
configurations separated by 200 time steps to make them independent.
All curves are for the same density $\rhobarm=0.050$
of the northbound particles and 
each curve is for a different value of
the boundary densities $\etabarp$ of the
eastbound particles.

Fig.\,\ref{fig_cylDthetacolumn} calls for several comments.
All curves show similar behavior:
as $i$ increases from 1 to $M=500$, the angle $\Delta\theta(i)$
first has a narrow `boundary' plateau,
then a rapid decrease, and then what seems like
a very wide and stable final `bulk' plateau.

The first plateau  corresponds to the boundary layer of width $\xi$,
here equal to $\xi\approx 75$
(if we take the point of reference in the zone
of rapid decrease at half the height difference between the two
plateaus), independently of the value of $\etabarm$. 
As discussed in section \ref{seccomparison}
the values of $\Delta\theta(i)$ obtained
in this boundary layer by straightforwardly applying Eq.\,(\ref{xtanthetavv}),
cannot be related to any angle of inclination.
We can however understand the value of the boundary plateau.
Near the west entrance the average density of the entering
eastbound particles should equal the value imposed by the boundary
condition, that is, $\la\rhop(1)\ra \simeq \etabarp$. 
When both species are uniformly distributed
(which corresponds to a disordered particle system),
the expected average speeds are $\vp=1-\rhobarm$ and $\vm=1-\etabarp$
so that for $i=1$ we expect 
\beq
\tan\Big( \frac{\pi}{4}+\Delta\theta(1) \Big)
= \frac{1-\etabarp}{1-\rhobarm}\,.
\label{xtanthetacyl1}
\eeq
Upon expanding to linear order in $\Delta\theta(1)$ we get
\beq
\Delta\theta(1) = \frac{\rhobarm-\etabarp}{2(1-\rhobarm)}\,.
\label{xDeltathetacyl1}
\eeq
This formula is satisfied quite well by the boundary plateau values 
in Fig.\,\ref{fig_cylDthetacolumn}.
Again, we repeat that $\Delta\theta_0$ does not here have the
interpretation of a slope of stripes.

\begin{figure}
\begin{center}
\scalebox{.35}
{\includegraphics{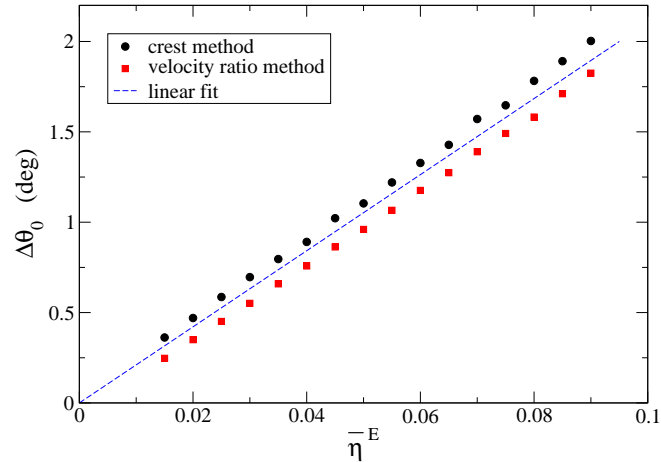}}
\end{center}
\caption{\small Chevron angle $\Delta\theta_0$ as a function of the
  average density $\etabar^{\pE}$ imposed at the open west boundary
  and for fixed $\rhobar^\pN=0.050$, obtained by simulation of the mean
  field equations (\ref{mfeqns2d}) and measured 
  by both the crest and the velocity ratio method, on a
  square lattice of linear size $M=500$ with cylindrical boundary conditions.
  The error bars are of the same order as the symbols.
  The straight line through the origin is the closest
  fit to both data sets; it has a slope of $21^{\circ}$. 
  } 
\label{fig_cylDthetafin}
\end{figure}

Of principal interest here, however, are the values of the bulk plateau.
These correspond to the bulk region on the cylinder surface, 
where we have
stripes with a single slope different from $45^\circ$, 
rather than chevrons.
We will nevertheless continue to speak of the `chevron effect' in this
case, too. Actually, in the bulk $\Delta\theta(i)$
seems to show a very slight
increase with $i$, as is clear in particular for the smaller values of
$\etabarp$. 
In order to arrive at a unique value for $\Delta\theta_0$ we 
determined $\Delta\theta_0$ as the
average of $|\Delta\theta(i)|$ over the columns 
with $200 \leq i \leq 300$, then averaged over at least 60 determinations.
The results are represented by the red square dots in
Fig.\,\ref{fig_cylDthetafin}.

Along with the velocity ratio method we applied to the same field 
configurations also the crest method. 
The results are represented by the black round dots in the same figure.

The error bars for each method are of the order of the symbol size;
they were estimated from variances
obtained by dividing the data for each data point into five subsets.
The results of the two methods are sufficiently close
that we may speak of a coherent picture.
They are nevertheless clearly distinct: the error bars do not overlap.
One factor that may contribute to this difference is the fact that
Eq.\,(\ref{xtantheta}) is valid only for stripes that are mutually
impenetrable, a condition that is not necessarily fully satisfied
in the model.

%%%%%%%%%%%%%%%%%%%%%%%%%%%%%%%%%%%%
%%%%%%%%%%%%%%%%%%%%%%%%%%%%%%%%%%%%
%%%%%%%%%%%%%%%%%%%%%%%%%%%%%%%%%%%%
%%%%%%%%%%%%%%%%%%%%%%%%%%%%%%%%%%%%

%%%%%%%%%%%%%%%%%%%%%%%%%%%%%%%%%%%%
%%%%%%%%%%%%%%%%%%%%%%%%%%%%%%%%%%%%
%%%%%%%%%%%%%%%%%%%%%%%%%%%%%%%%%%%%
%%%%%%%%%%%%%%%%%%%%%%%%%%%%%%%%%%%%

%%%%%%%%%%%%%%%%%%%%%%%%%%%%%%%%%%%%
%%%%%%%%%%%%%%%%%%%%%%%%%%%%%%%%%%%%
\begin{figure}
\begin{center}
\scalebox{.35}
{\includegraphics{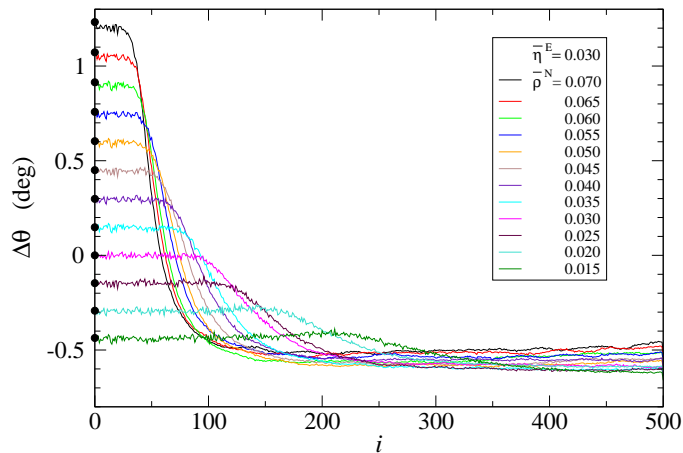}}
\end{center}
\caption{\small Chevron angle $\Delta\theta$ as a function of the
  column index $i$,
  obtained by simulation of the mean field equations (\ref{mfeqns2d}) 
  and measured by means of the velocity ratio method, on a
  square lattice of linear size $M=500$ with cylindrical boundary
  conditions. The eastbound particles have a fixed boundary density
  $\etabarp=0.030$. Curves for various densities $\rhobarm$
  of the northbound particles show different penetration depths $\xi$ but
  have closely similar plateau values for $i \gtrsim \xi$.
  The dots for $i=1$ are the boundary values predicted by 
  Eq.\,(\ref{xDeltathetacyl1}).
  } 
\label{fig_cylDthetarhop}
\end{figure}
%%%%%%%%%%%%%%%%%%%%%%%%%%%%%%%%%%%%
%%%%%%%%%%%%%%%%%%%%%%%%%%%%%%%%%%%%

Fig.\,\ref{fig_cylDthetarhop} shows another set of curves of the 
column dependent chevron angle, obtained in the same way as those of
Fig.\,\ref{fig_cylDthetacolumn}, but now all curves are
for the same $\etabar^{\pE}=0.030$ and each one is for a different
value of $\rhobarm$. 
It appears that all these curves
have plateau values $-\Delta\theta_0$ with 
$0.50^{\circ}\lesssim\Delta\theta_0\lesssim 0.60^{\circ}$.
which is fully compatible with 
the data point of Fig.\,\ref{fig_cylDthetafin} 
for $\etabar^{\pE}=0.030$, namely 
$\Delta\theta_0=0.55^{\circ}\pm0.02^{\circ}$.
There is however a slight drift of the plateau value with increasing $i$.
This effect becomes more pronounced as $\etabarp$ gets larger but we have
not pursued our investigation of this point.
We notice that
the values $\Delta\theta(1)$ are again in perfect agreement
with Eq.\,(\ref{xDeltathetacyl1}).
Finally, Fig.\,\ref{fig_cylDthetarhop} shows the variation of the
width of the boundary layer with $\rhobarm$. 
If we let again the points at mid-height between the boundary plateau
and the bulk plateau 
determine the penetration depth $\xi$, a crude fit
shows that $\xi\approx 4.2/\rhobarm$ for $\rhobarm\to 0$.
\vspace{2mm}

The cylinder study described here was
carried out for the mean-field model of Eqs.\,(\ref{mfeqns2d}).
Similar results for the particle model, not reported here, 
show that in that system, too, the chevron angle is linear in the density
of the eastbound particles and independent of the density of the
northbound ones. In each case the explanation lies in the 
asymmetry caused by the fact that the entering
particle species is fully disordered whereas the other species
has had time to organize.

These density dependencies are the main result of our investigation
with CBC boundary conditions.

%%%%%%%%%%%%%%%%%%%%%%%%%%%%%%%%%%%%%%%%%%%%%%%%%%%%%%%%%%%%%%%%%%%%%%%%%%%%%

\subsection{Chevron effect in a transient}
\label{secchevrontransient}

The system to be studied in this final section has been designed 
for the sole purpose of testing our understanding of the chevron effect.
Whereas until now we dealt exclusively with stationary state properties,
we will here consider a {\it transient\,} effect,
and that for a very particular set of initial conditions.
We consider again the mean field equations (\ref{mfeqns2d})
in cylindrical geometry and
prepare the system at time $t=0$ in a state with the uniform nonrandom 
initial condition $\rhop_0(\brr)=\rhom_0(\brr)=\rhobar_0$.
This initial state would be stationary if there were no open boundaries.
We however evolve this system in time 
with the same random density boundary condition 
as before along the west entrance, characterized by an $\etabarp$, and with free exit at the east boundary.

\begin{figure}
\begin{center}
\scalebox{.4}
{\includegraphics{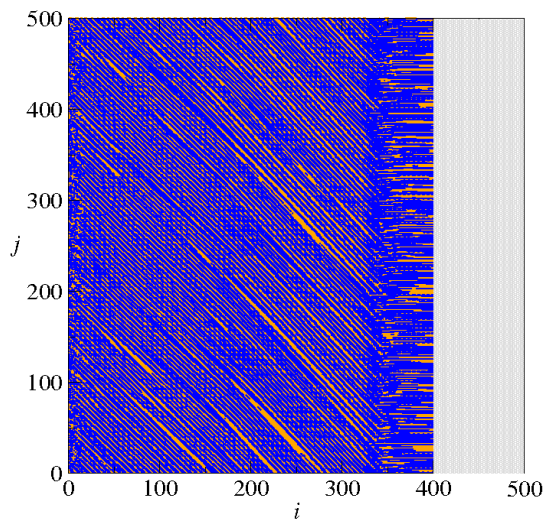}} 
\end{center}
\caption{\small Density field showing the transient configuration
at time $t=400$ on an intersection square of linear size $M=500$
with cylindrical boundary conditions. The color code is as in
Fig.\,\ref{fig_NLfield}. The west entrance boundary
is subject to a random boundary density of average $\etabarp=0.10$. 
The nonrandom initial condition was $\rhopm_0(\brr)=0.10$.
In the grey area these initial values still persist at time $t=400$.
}
\label{fig_transdensity}
\end{figure}

\begin{figure}
\begin{center}
\scalebox{.4}
{\includegraphics{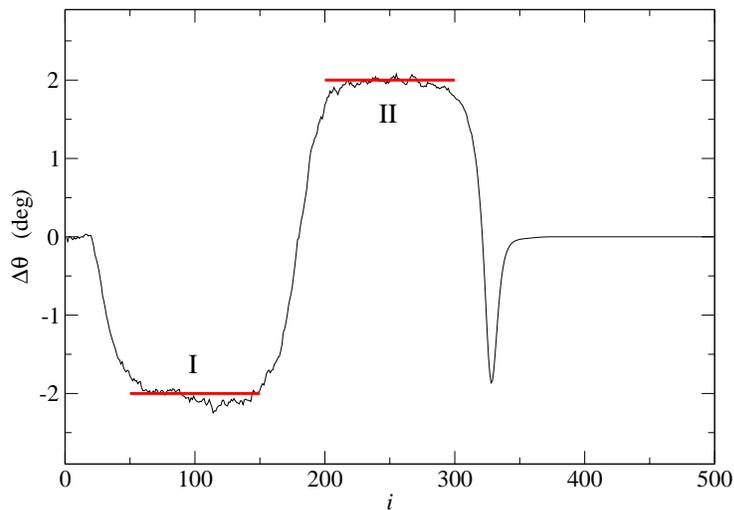}} 
\end{center}
\caption{\small Chevron angle $\Delta\theta$
as a function of the column index $i$,
determined by the velocity ratio method
in the configuration of Fig.\,\ref{fig_transdensity}.
The heavy horizontal red lines mark the plateau values $\pm 2^{\circ}$.
}
\label{fig_transDtheta}
\end{figure}

We considered specifically a system of linear size $M=500$ having
$\rhobarm=\etabarp=\rhobar_0=0.10$.
At time $t=400$, the stationary state has not set in yet. 
A density plot of the fields then looks like in
Fig.\,\ref{fig_transdensity}.
In Fig.\,\ref{fig_transDtheta} we show the corresponding column dependent 
values of the chevron angle $\Delta\theta(i)$. These figures call for
the following comments.

We now discuss Fig.\,\ref{fig_transdensity} and Fig.\,\ref{fig_transDtheta} in the order of decreasing column index.
Since the influence from the boundary penetrates into the bulk 
by one lattice unit per time step, at time $t=400$ the region of the
intersection square with column index $i\geq 400$, colored gray
in the figure, has remained in the initial uniform state.
In the region $325\lesssim i<400$
the amplitude of the
perturbation decays exponentially;
we will discuss this zone in greater detail in another article
\cite{cividini_h2013}.

Of main interest is the region $50\lesssim i\lesssim325$, 
which has the diagonally striped structure
characteristic of the crossing flows.
The region consists of two zones, I and II, extending between
$50\lesssim i\lesssim 175$ and $200\lesssim i\lesssim 325$,
respectively. 
Although barely visible in Fig.\,\ref{fig_transdensity}, 
the stripes in the 
zones I and II have different angles of inclination $\theta$.
This becomes very clear in Fig.\,\ref{fig_transDtheta}, which shows 
the chevron angle $\Delta\theta(i)=\theta(i)-\frac{\pi}{4}$
as a function of the column index $i$.
In zones I and II this angle has two distinct plateau values
close to $\mp\Delta\theta_0$, respectively, where
$\Delta\theta_0=2.0^{\circ}$. 
The two zones are separated by a transition layer
and zone I is separated from the boundary by the usual boundary layer.

During the time evolution the widths of zones I and II increase roughly
linearly with $t$ whereas the transition layer
keeps a constant width. Hence zone I gradually extends all the way
to the east end of the interaction square and forces zone II 
(and with it one leg of the chevron) out of the system. 
What remains is a stationary state of the type studied 
in section \ref{secchevronstationary}, with a value
$-\Delta\theta_0=-2.0^{\circ}$ for the chevron angle.
\footnote{
This value, for a system with imposed boundary density $\etabarp=0.10$, 
is fully consistent with the
data set of the red squares, when slightly extrapolated, of
Fig.\,\ref{fig_cylDthetafin}.} 

The appearance during the transient of a zone II with the opposite value
of the chevron angle needs to be explained.
The explanation follows from rule \ref{ruleangle} (section \ref{section:chevrontheory}).
In the present case, at the right hand interface of zone II
the northbound particles constitute the disordered species which
invades the eastbound ones that are diagonally ordered.
Therefore, according to the rule, 
the angle between the diagonals and the direction of
propagation of the disordered species (which here moves northward)
is reduced by $\frac{1}{2}\rhobarm$; 
and Fig.\,\ref{fig_transDtheta} shows exactly that effect.
The difference with what happens at the entrance boundary is that here
the ordered species moves perpendicular to the interface and the
disordered one parallel to it.

%%%%%%%%%%%%%%%%%%%%%%%%%%%%%%%%%%%
%%%%%%%%%%%%%%%%%%%%%%%%%%%%%%%%%%%
%%%%%%%%%%%%%%%%%%%%%%%%%%%%%%%%%%%
%%%%%%%%%%%%%%%%%%%%%%%%%%%%%%%%%%%

%%%%%%%%%%%%%%%%%%%%%%%%%%%%%%%%%%%
%%%%%%%%%%%%%%%%%%%%%%%%%%%%%%%%%%%
%%%%%%%%%%%%%%%%%%%%%%%%%%%%%%%%%%%
%%%%%%%%%%%%%%%%%%%%%%%%%%%%%%%%%%%

%%%%%%%%%%%%%%%%%%%%%%%%%%%%%%%%%%%%%%%%%%%%%%%%%%%%%%%%%%%%%%%%%%%%%%%%%%%%%

%%%%%%%%%%%%%%%%%%%%%%%%%%%%%%%%%%%%%%%%%%%%%%%%%%%%%%%%%%%%%%%%%%%%%%%%%%%%%%%

\section{Summary and conclusion}
\label{section:conclusion}

We have considered in this work a class of theoretical
models of two crossing unidirectional traffic flows, 
one composed of eastward and one of northward traveling particles. 
The basic model parameters are the street width $M$ 
and either the imposed current or the imposed particle density.
Because of their simplicity, we believe that this class of models
has an intrinsic
interest as an example of a driven nonequilibrium system.

We have discussed a phenomenon observed widely
in more realistic many-parameter models as well as in experiments \cite{hoogendoorn_d2003}, namely
the instability -- in the crossing area -- 
of the randomly uniform state against
segregation into diagonal stripes of alternatingly northward and
eastward traveling species. We have shown that during the
development of the instability the particles of
each species aggregate into string-like structures.

The principal models in our class are
a particle model with two different update rules
and a closely related mean field model.
The latter has allowed us to provide
an analytic explanation of the instability in the
simplest possible context,
namely when the two flows go around a torus.
Such toroidal geometry has become popular since the introduction
of the BML model \cite{biham_m_l1992}. 
The linear stability analysis that we performed for the torus may
be extended to the open intersection square; that calculation
is however very cumbersome and will be the subject of a future
publication \cite{cividini_h2013}.

We have moreover discovered that for a crossing with open boundary
conditions, 
which is the case of principal interest in this work,
the stripes actually have two branches that join to form a chevron.
The slopes of the branches (with respect to the two flow directions) 
differ from
$45^{\circ}$ by an amount $\pm\Delta\theta_0$ that we call the 
{\it chevron angle}. 
The angle is negative in the upper triangular half of the
intersection square and positive in the lower half.
Its absolute value is very small (less than $2^{\circ}$ in all cases
studied), but the chevron effect is robustly present
in all model versions that we studied.
We found by simulation that $\Delta\theta_0$ is linear
in the density of the particles coming in through a boundary, and
independent of the density of the particles moving parallel to that
boundary.

The chevron phenomenon disappears in the limit of zero particle density
for two different reasons.
First of all, the chevron angle becomes small,
and secondly, the boundary layer of width $\xi$ beyond which it is visible, 
extends further and further into the system.
So we cannot study the chevron effect
in the limit $\rho\to 0$ in finite systems,
and in this sense the effect is nonperturbative.
\vspace{2mm}

In section \ref{section:chevrontheory} we provided some elementary theoretical
arguments that explain the chevron effect and that involve
the formation of linear aggregates (stripes) of same-type particles. 
We obtained an approximate but explicit formula for the
chevron angle as a function of the particle density.

The theory is based on the special limiting situation in
which one particle type is fully aggregated into strings
and the other one randomly and uniformly 
distributed in space, an asymmetry indeed clearly observed in the
simulations. 
One may nevertheless consider that
the theory of the chevron effect still needs further development.
It is tempting to speculate that there exists 
a hydrodynamic theory with two
components each of which is characterized not only by its average
local density and velocity, but also by a variable expressing its
degree of aggregation. 
In future work
\cite{cividini_a2013} we will return to related questions
and study the interaction between two particles traveling 
on parallel lanes,  
as mediated by a sea of perpendicular particles.
\vspace{2mm}

The various different manifestations of the stripe formation
instability and the chevron effect studied above
point to the conclusion that these phenomena occur generically
whenever we have crossing unidirectional flows
with hard core interaction and deterministic rules of motion.

Further questions that may be asked concern modifications
of the model. 
What happens, for example, if blocked particles are allowed to jump
laterally? What happens if, for open boundary conditions, the two
directions have different imposed flow rates?
Furthermore,
from a theoretical point of view the $M\to\infty$ limit of this system
is interesting.
As far as we have been able to ascertain, the `chevron' states that we
discovered are stable stationary states, but we are not sure what they
become in the infinite system limit.
A related question concerns
the behavior in an anisotropic geometry,
that is, in an $M_1\times M_2$ intersection rectangle where
one side of the rectangle might tend to infinity with the other one fixed.

Some of these questions will be the subject of future work.

%%%%%%%%%%%%%%%%%%%%%%%%%%%%%%%%%%%%%%%%%%%%%%%%%%%%%%%%%%%%%%%%%%%%%%%%%%%%%%%

\section*{Acknowledgments}

We thank R.K.P.\,Zia for useful discussions.


\vspace{10mm}

\begin{thebibliography}{10}
\expandafter\ifx\csname url\endcsname\relax
  \def\url#1{\texttt{#1}}\fi
\expandafter\ifx\csname urlprefix\endcsname\relax\def\urlprefix{URL }\fi

\bibitem{schadschneider2008b}
A.~Schadschneider, Modelling of transport and traffic problems, Lecture Notes
  in Computer Science 5191 (2008) 22--31.

\bibitem{helbing2001b}
D.~Helbing, Traffic and related self-driven many-particle systems, Reviews of
  Modern Physics 73 (2001) 1067--1141.

\bibitem{cross_h1993}
M.~C. Cross, P.~C. Hohenberg, Pattern formation outside of equilibrium, Rev.
  Mod. Phys. 65 (1993) 851--1108.

\bibitem{wolfram1983}
S.~Wolfram, Statistical mechanics of cellular automata, Rev. Mod. Phys. 55
  (1983) 601--644.

\bibitem{packard_w1985}
N.~Packard, S.~Wolfram, Two-dimensional cellular automata., J. Stat. Phys. 38
  (1985) 901--946.

\bibitem{wolfram1984}
S.~Wolfram, Universality and complexity in cellular automata., Physica D 10
  (1984) 1--35.

\bibitem{nagel_s1992}
K.~Nagel, M.~Schreckenberg, A cellular automaton model for freeway traffic, J.
  Physique I 2 (1992) 2221--2229.

\bibitem{fouladvand_s_s04b}
M.~E. Fouladvand, Z.~Sadjadi, M.~R. Shaebani, Optimized traffic flow at a
  single intersection: traffic responsive signalization, J. Phys. A: Math. Gen.
  37 (2004) 561--576.

\bibitem{foulaadvand_n07}
M.~E. Foulaadvand, M.~Neek-Amal, Asymmetric simple exclusion process describing
  conflicting traffic flows, EPL 80.

\bibitem{du2010}
H.-F. Du, Y.-M. Yuan, M.-B. Hu, R.~Wang, R.~Jiang, Q.-S. Wu, Totally asymmetric
  exclusion processes on two intersected lattices with open and periodic
  boundaries, J. Stat. Mech. (2010) P03014.

\bibitem{appert-rolland_c_h2011c}
C.Appert-Rolland, J.Cividini, H.J.Hilhorst, Intersection of two tasep traffic
  lanes with frozen shuffle update, J. Stat. Mech. (2011) P10014.

\bibitem{biham_m_l1992}
O.~Biham, A.~Middleton, D.~Levine, Self-organization and a dynamic transition
  in traffic-flow models, Phys. Rev. A 46 (1992) R6124--R6127.

\bibitem{ding_j_w2011}
Z.-J. Ding, R.~Jiang, B.-H. Wang, Traffic flow in the biham-middleton-levine
  model with random update rule, Phys. Rev. E 83 (2011) 047101.

\bibitem{ding2011}
Z.-J. Ding, R.~Jiang, W.~Huang, B.-H. Wang, Effect of randomization in the
  biham–middleton–levine traffic flow model, J. Stat. Mech. (2011) P06017.

\bibitem{hoogendoorn_b2003}
S.~Hoogendoorn, P.~H. Bovy, Simulation of pedestrian flows by optimal control
  and differential games, Optim. Control Appl. Meth. 24 (2003) 153--172.

\bibitem{moussaid2012}
M.~Moussaïd, E.~Guillot, M.~Moreau, J.~Fehrenbach, O.~Chabiron, S.~Lemercier,
  J.~Pettré, C.~Appert-Rolland, P.~Degond, G.~Theraulaz, Traffic instabilities
  in self-organized pedestrian crowds, PLoS Computational Biology 8 (2012)
  1002442.

\bibitem{schmittmann_z1995}
B.~Schmittmann, R.~Zia, Statistical Mechanics of driven diffusive systems,
  Vol.~17 of Phase Transitions and Critical Phenomena, Academic Press, New
  York, 2013.

\bibitem{dzubiella_h_l2002}
J.~Dzubiella, G.~P. Hoffmann, H.~L{\"o}wen, Lane formation in colloidal
  mixtures driven by an external field, Phys. Rev. E 65 (2002) 021402.

\bibitem{yamamoto_o2011}
K.~Yamamoto, M.~Okada, Continuum model of crossing pedestrian flows and swarm
  control based on temporal/spatial frequency, in: 2011 IEEE International
  Conference on Robotics and Automation, 2011.

\bibitem{hilhorst_a2012}
H.~Hilhorst, C.~Appert-Rolland, A multi-lane {TASEP} model for crossing
  pedestrian traffic flows, J. Stat. Mech. (2012) P06009.

\bibitem{cividini_a_h2013}
J.~Cividini, C.~Appert-Rolland, H.~Hilhorst, Diagonal patterns and chevron
  effect in intersecting traffic flows, Europhys. Lett. 102 (2013) 20002.

\bibitem{appert-rolland_c_h2011a}
C.Appert-Rolland, J.~Cividini, H.J.Hilhorst, Frozen shuffle update for an
  asymmetric exclusion process on a ring, J. Stat. Mech. (2011) P07009.

\bibitem{appert-rolland_c_h2011b}
C.Appert-Rolland, J.Cividini, H.J.Hilhorst, Frozen shuffle update for an
  asymmetric exclusion process with open boundary conditions, J. Stat. Mech.
  (2011) P10013.

\bibitem{cividini_h2013}
J.~Cividini, H.~Hilhorst, In preparation.

\bibitem{cividini_a2013}
J.~Cividini, C.~Appert-Rolland, Wake-mediated interaction between driven
  particles crossing a perpendicular flow, arXiv:1305.3206.

\bibitem{hoogendoorn_d2003}
S.~P. Hoogendoorn, W.~Daamen, Self-organization in walker experiments, in:
  S.~Hoogendoorn, S.~Luding, P.~Bovy, et~al. (Eds.), Traffic and Granular Flow
  '03, Springer, 2005, p.~??

\end{thebibliography}
\end{document}